\documentclass[english,journal=nalefd,layout=twocolumn,manuscript=letter]{achemso}
\usepackage[T1]{fontenc}
\usepackage[latin9]{inputenc}
\usepackage{color}
\usepackage{babel}
\usepackage{textcomp}
\usepackage{amsmath}
\usepackage{amssymb}
\usepackage{graphicx}
\usepackage{hyperref}
 
\makeatletter

\title{Polaritonic Bottleneck in Colloidal Quantum Dots}

\author{Kaiyue Peng}

\affiliation{Department of Chemistry, University of California, Berkeley, California
94720, United States}

\email{kaiyue_peng@berkeley.edu}

\author{Eran Rabani}

\affiliation{Department of Chemistry, University of California, Berkeley, California
94720, United States}

\altaffiliation{Materials Sciences Division, Lawrence Berkeley National Laboratory,
Berkeley, California 94720, United States}

\alsoaffiliation{The Sackler Center for Computational Molecular and Materials Science,
Tel Aviv University, Tel Aviv, Israel 69978}

\email{eran.rabani@berkeley.edu}


\setkeys{acs}{articletitle = true}
\usepackage{bm}
\AbstractOn
\SectionsOn

\makeatother

\begin{document}
\begin{abstract}
Controlling the relaxation dynamics of excitons is key to improving the efficiencies of semiconductor--based applications. Confined semiconductor nanocrystals (NCs) offer additional handles to control the properties of excitons, for example, by changing their size or shape, resulting in a mismatch between excitonic gaps and phonon frequencies. This has led to the hypothesis of a significant slowing--down of exciton relaxation in strongly confined NCs, but in practice due to increasing exciton--phonon coupling and rapid multiphonon relaxation channels, the exciton relaxation depends only weakly on the size or shape. Here, we focus on elucidating the nonradiative relaxation of excitons in NCs placed in an optical cavity. We find that multiphonon emission of carrier governs the decay resulting in a polariton--induced phonon bottleneck with relaxation timescales that are slower by orders of magnitude compared to the cavity--free case, while the photon fraction plays a secondary role.
\end{abstract}

The electronic and optical properties of semiconductor quantum dot (QD) nanocrystals (NCs) have been widely studied over the past decades, leading to the development of colloidal--based optoelectronic devices and remarkable advancements in new commercialized technologies.\cite{hou_incoherent_2023,bhattacharya2004quantum,hanifi2019redefining} A key to the rational design of NC--based technologies with increased quantum yields and reduced thermal losses relies on controlling the radiative, nonradiative, dephasing, and energy transfer channels, governed by exciton--exciton and exciton--phonon couplings.\cite{philbin2020auger,jasrasaria2022simulations} One of the most important relaxation channels limiting the efficiency of NC--based devices involves the nonradiative relaxation of excitons to form a band edge excitation. It has been argued that due to a mismatch between the excitonic energy gaps and the lattice vibrations, the relaxation of excitons in strongly confined NCs would be extremely slow, resulting in a "phonon bottleneck", despite enhancements of the exciton--phonon couplings.\cite{melnychuk2021multicarrier} However, measurements of the cooling process in NCs have yielded conflicting results, where most time--resolved measurements do not support the existence of a phonon bottleneck and the relaxation of excitons occurs on sub--picosecond timescales for various NC materials and sizes.\cite{schaller2005breaking,cooney2007breaking,cooney2007unified} This unexpected result was recently rationalized theoretically signifying the role of multiphonon processes in NCs, resulting in ultrafast exciton relaxation to the band edge.\cite{jasrasaria2023circumventing}

\begin{figure}[t]
\begin{centering}
\includegraphics[width=7.5cm]{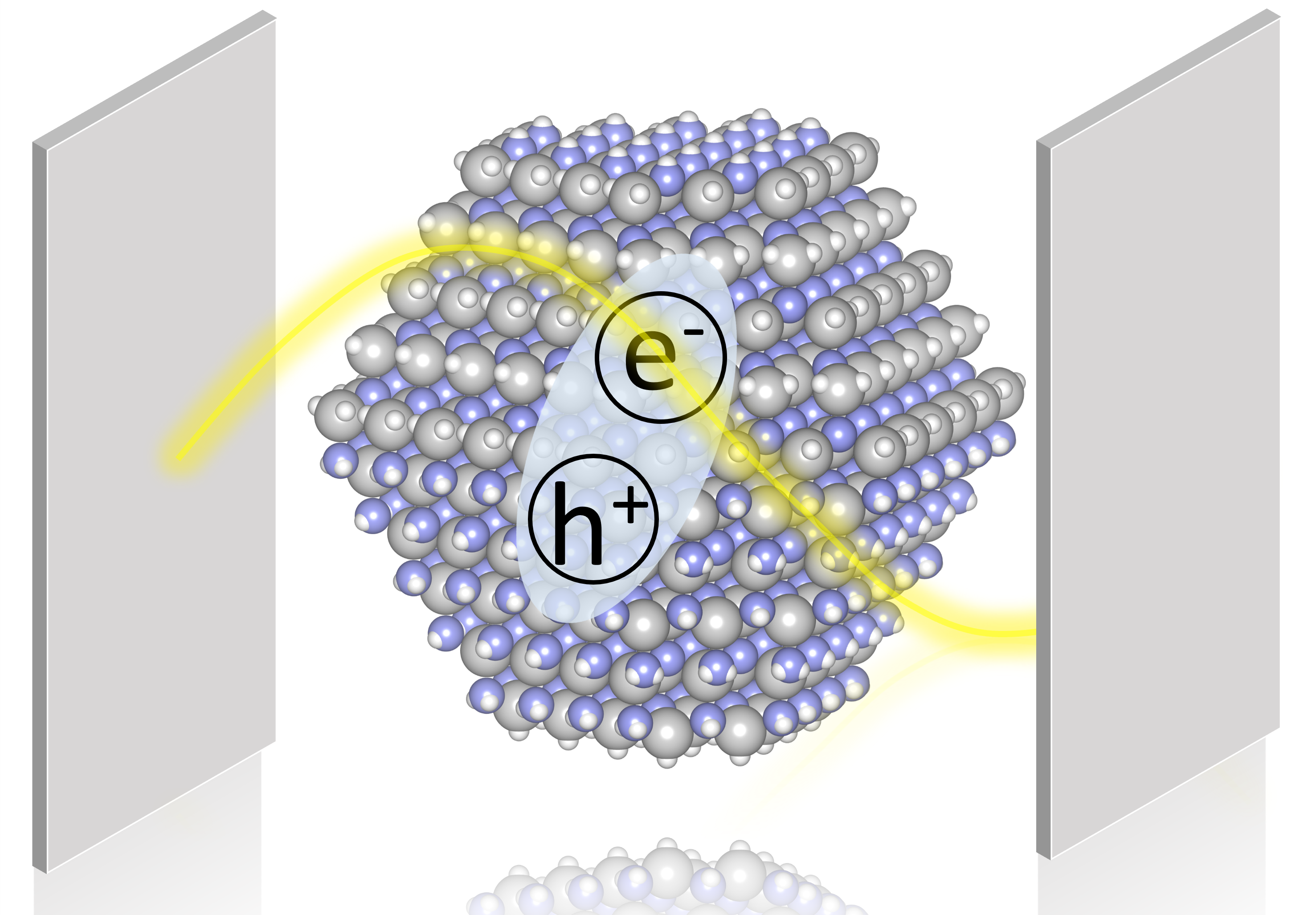}
\par\end{centering}
\caption{An illustration of a nanocrystal QD in an optical cavity.\label{fig:illustration}}
\end{figure}

Controlling the nonradiative relaxation timescales and relaxation pathways of excitons has been central for improved quantum efficiencies.\cite{hu2022tuning,fomenko2008solution,talapin2010prospects} The use of heterostructure core--shell NCs, for example, results in slower relaxation rates compared to core NCs, mainly due to the reduction of the couplings between excitons and surface phonon modes,\cite{jasrasaria2021interplay} with relaxation rates that are slower by a factor of $5$.\cite{jasrasaria2023circumventing} An alternative approach to modify the electronic and vibronic properties of the NC is by strongly coupling to light, for example, inside a cavity (see Fig.~\ref{fig:illustration} for an illustration). The properties of such hybrid exciton--photon states (polaritonic states) can be tuned without changing the size and composition of the NC, by modulating the cavity parameters such as the cavity photon energy and/or the coupling strength between excitons and photons.\cite{schafer2019modification,du2018theory,gonzalez2015harvesting,westmoreland2019properties} 

This approach has been applied for J--aggregates, two--dimensional quantum wells, and bulk materials,\cite{coles2013imaging,laitz2023uncovering,michetti2008simulation,stokker2008polariton,du2018theory} and is also popular for controlling the electronic properties of molecular systems, and in particular how the products in chemical reactions can be tuned by coupling with light, with reported reactions being selective, slower or faster in optical cavities.\cite{mandal2019investigating,schafer2022shining} In addition, such hybrid states exhibit unique properties that make them relevant for a wide range of applications, including lasers,\cite{kena2010room,mazza2013microscopic,fraser2016physics,imamog1996nonequilibrium,bajoni2012polariton} Bose--Einstein condensates,\cite{amo2009collective,kasprzak2006bose,deng2010exciton} and quantum bits.\cite{ghosh2020quantum,xue2021split,demirchyan2014qubits} A deep understanding along side a theoretical description of the polaritonic dynamics is available for molecules and relatively small system sizes, while the description of realistic materials is still lacking.

In this work, we consider a CdSe NC in a cavity and study the relaxation pathways and timescales of polaritonic states as we tune the cavity photon energy ($\hbar\omega_{c}$) and cavity--exciton coupling ($\hbar g$).  We parametrize the Pauli--Fierz Hamiltonian\cite{mandal_theoretical_2023} to describe a lossless micro--cavity confined single photon mode coupled to a manifold of excitonic states. The latter were calculated within the atomistic semiempirical pseudopotential model combined with the Bethe--Salpeter equation (BSE), accounting for electron--hole correlations.\cite{jasrasaria2021interplay}  We perform a small polaron transformation~\cite{jasrasaria2023circumventing,xu2016non} on the Pauli--Fierz Hamiltonian to rescale the polariton--phonon coupling (obtained from the semiempirical pseudopotential model) and use a quantum master equation with memory approximated to second--order in this coupling (Redfield equation). This transformation allows us to account for multiphonon processes within the lowest--order perturbation scheme. 

We find that the relaxation to thermal equilibrium is hindered by the coupling to the cavity mode, which can be controlled by changing the micro--cavity photon energy or by changing the cavity--exciton coupling. We explore the role of temperature and cavity parameters on the timescales of relaxation and find that as the cavity photon energy decreases and/or the cavity--exciton coupling increases, the relaxation to thermal equilibrium is significantly slowed down. This polaritonic--induced phonon bottleneck is attributed to the increase in the polaritonic gap of the two lowest polaritonic states and to changes in the polariton--phonon coupling strength.

We adopt the following model Hamiltonian to describe the ground state and a manifold of excitonic states coupled to the vibrational modes of a nanocrystal QD, with exciton--phonon coupling (EXPC) expanded to the lowest order in the vibrational normal modes:\cite{jasrasaria2021interplay,jasrasaria2023circumventing}

\begin{align}
H_{\text{QD}} & =E_{g}\left|\psi_{g}\right\rangle \left\langle \psi_{g}\right|+\sum_{n}E_{n}\left|\psi_{n}\right\rangle \left\langle \psi_{n}\right|+\nonumber \\
+ & \sum_{\alpha}\hbar\omega_{\alpha}b_{\alpha}^{\text{\ensuremath{\dagger}}}b_{\alpha}+\sum_{nm\alpha}V_{nm}^{\alpha}\left|\psi_{m}\right\rangle \left\langle \psi_{n}\right|q_{\alpha}.\label{eq:hamil-excitons}
\end{align}
In the above equation, $E_{g}$ is the energy of the ground state
$\left|\psi_g\right\rangle $ and $E_{n}$ is the energy of an exciton in state $\left|\psi_{n}\right\rangle $. The phonon frequencies $\omega_{\alpha}$ and phonon modes $q_{\alpha}=\sqrt{\frac{\hbar}{2\omega_{\alpha}}}\left(b_{\alpha}^{\text{\ensuremath{\dagger}}}+b_{\alpha}\right)$ were obtained by diagonalizing the dynamical matrix of the NC calculated using the Stillinger--Weber force field.\cite{zhou2013stillinger} The excitonic states and the EXPC elements, $V_{nm}^{\alpha}$, were
calculated using the semiempirical pseudopotential methods~\cite{rabani1999electronic} combined with the BSE~\cite{rohlfing2000electron}, including only spin--allowed bright or dim states.\cite{jasrasaria2021interplay,jasrasaria2022simulations} We ignore the EXPC between the ground state and all the other excitonic
states, which would lead to nonradiative relaxation of excitons to the ground state on timescales much longer than the current interest. This parametrization procedure was validated against optical stokes shift measurements,\cite{jasrasaria2021interplay} single--particle photoluminescence lineshape broadening,\cite{doi:10.1021/acs.jpclett.3c01630}
and most recently to the relaxation dynamics of excitons in both core and core/shell NCs.\cite{brosseau2023ultrafast,jasrasaria2023circumventing}

The nanocrystal QD is placed in an optical cavity modeled by a single--mode Pauli\textminus Fierz Hamiltonian:\cite{mandal2020polariton}
\begin{align}
H_{\text{cav}} & =\hbar\omega_{c}a^{\dagger}a+\sum_{n}\hbar g_{n}\left(a^{\dagger}\left|\psi_g\right\rangle \left\langle \psi_{n}\right|+h.c.\right)\nonumber \\
+ & \frac{\hbar}{\omega_{c}}\left[\text{\ensuremath{\sum_{n}g_{n}^{2}}}\left|\psi_g\right\rangle \left\langle \psi_{g}\right|+\sum_{nm}g_{n}g_{m}\left|\psi_{n}\right\rangle \left\langle \psi_{m}\right|\right],\label{eq:hamil-cav}
\end{align}
where $a$ ($a^{\dagger}$) is the destruction (creation) operator
of a photon with frequency $\omega_{c}$ and $g_{n}$ is coupling
strength between the cavity mode and an exciton in state $\left|\psi_{n}\right\rangle $.
The latter is given by $g_{n}=\sqrt{\frac{\omega_{c}}{2\hbar\varepsilon\mathcal{V}}}\boldsymbol{\mu}_{n}\cdot\boldsymbol{k}$,
where $\boldsymbol{\mu}_{n}=\left\langle \psi_{n}\right|\hat{\boldsymbol{\mu}}\left|\psi_g\right\rangle $
is the transition dipole moment from the ground state to exciton $\left|\psi_{n}\right\rangle $
(obtained from the BSE calculation~\cite{philbin2018electron}) and $\boldsymbol{k}$ is a unit vector of the polarization direction of the electromagnetic field. Finally, $\varepsilon$ is the effective permittivity inside the cavity, and $\mathcal{V}$ is the effective cavity quantization volume.

The total Hamiltonian of the nanocrystal QD coupled to the cavity mode can be represented using the excitonic--photonic basis set, namely, $\left\{ \left|\psi_g;1\right\rangle ,\left|\psi_{n};0\right\rangle \right\} $, where $0,1$ represent the number of photons (we limit the discussion
to single--excitation manifold only). In this basis, the total Hamiltonian can be written as a sum of two terms, $H_{\text{PF}}=H_{\text{QD}}+H_{\text{cav}}=H_{\text{S}}+H_{\text{ph}}$,
where
\begin{align}
H_{\text{S}} & =\left(\hbar\omega_{c}+\sum_{n}\frac{\hbar}{\omega_{c}}g_{n}^{2}\right)\left|\psi_g,1\right\rangle \left\langle \psi_{g},1\right|+\nonumber \\
+ & \sum_{n}E_{n}\left|\psi_{n},0\right\rangle \left\langle \psi_{n},0\right|+\nonumber \\ 
+ & \frac{\hbar}{\omega_{c}}\sum_{nm}g_{n}g_{m}\left|\psi_{n},0\right\rangle \left\langle \psi_{m},0\right|\nonumber \\
+ & \sum_{n}\hbar g_{n}\left(\left|\psi_{n},0\right\rangle \left\langle \psi_{g},1\right|+\left|\psi_g,1\right\rangle \left\langle \psi_{n},0\right|\right),\label{eq:hamil_S}
\end{align}
and as before, the Hamiltonian describing the phonon modes is given by:
\begin{equation}
H_{\text{ph }}=\sum_{\alpha}\hbar\omega_{\alpha}b_{\alpha}^{\text{\ensuremath{\dagger}}}b_{\alpha}+\sum_{nm\alpha}V_{mn}^{\alpha}\left|\psi_{m},0\right\rangle \left\langle \psi_{n},0\right|q_{\alpha}.\label{eq:hamil_ph}
\end{equation}
The hybrid eigenstates of the $H_{\text{S}}$ are called polaritonic states and can be obtained by diagonalizing $H_{\text{S}}$. The total Hamiltonian ($\tilde{H}_\text{PF}$) is then given in terms of the polaritonic states, $\left|\varphi_{n}\right\rangle $, as:
\begin{align}
\tilde{H}_\text{PF} & =\sum_{n}\tilde{E}_{n}\left|\varphi_{n}\right\rangle \left\langle \varphi_{n}\right|+\sum_{\alpha}\hbar\omega_{\alpha}b_{\alpha}^{\text{\ensuremath{\dagger}}}b_{\alpha}\nonumber \\
+ & \sum_{mn\alpha}\tilde{V}_{mn}^{\alpha}\left|\varphi_{m}\right\rangle \left\langle \varphi_{n}\right|q_{\alpha},\label{eq:hamil-total}
\end{align}
where $\left|\varphi_{n}\right\rangle =c_{ng}\left|\psi_g,1\right\rangle +\sum_{m}c_{nm}\left|\psi_{m},0\right\rangle $
is a polaritonic state with energy $\tilde{E}_{n}$. $\tilde{V}_{mn}^{\alpha}$ represents the coupling matrix element between two polaritonic states
$\left|\varphi_{m}\right\rangle $ and $\left|\varphi_{n}\right\rangle $ via phonon mode $\alpha$. 

The above Hamiltonian has an identical form to that of a bare NC Hamiltonian, with the additional complexity of describing the dressed ground state ($\left|\psi_g,1\right\rangle $). To describe the relaxation dynamics of the polaritonic states generated by this Hamiltonian, we followed the procedure developed in Ref.~\citenum{jasrasaria2023circumventing}.
The first step involves a polaron transformation, which not only rescales the coupling between the polaritonic states and the phonon modes, but also allows for the description of multi--phonon relaxation within the lowest--order perturbation theory.
Such multi--phonon relaxation processes are central in describing the ultrafast dynamics of hot excitons.
The polaron transformation of the polaritonic Hamiltonian is similar to the one used for excitons, namely, ${\cal H}=e^{S}\tilde{H}_\text{PF}e^{-S}$,
where $S=-\sum_{\alpha}\frac{i\sum_{n}\tilde{V}_{nn}^{\alpha}p_{\alpha}}{\hbar\omega_{\alpha}^{2}}\left|\varphi_{n}\right\rangle \left\langle \varphi_{n}\right|$
and $p_{\alpha}$ is the momentum operator of vibrational mode $\alpha$.
The total transformed Hamiltonian is then given by: 
\begin{align}
{\cal H} & =\sum_{n}\left(\tilde{E}_{n}-\lambda_{n}\right)\left|\varphi_{n}\right\rangle \left\langle \varphi_{n}\right|+\sum_{\alpha}\hbar\omega_{\alpha}b_{\alpha}^{\text{\ensuremath{\dagger}}}b_{\alpha}\nonumber \\
+ & \sum_{n\neq m}\left(\sum_{\alpha}W_{nm}^{\alpha}q_{\alpha}-\lambda_{nm}\right)\left|\varphi_{n}\right\rangle \left\langle \varphi_{m}\right|\label{eq:Hamil-total-polaron}
\end{align}
where $\lambda_{n}=\frac{1}{2}\sum_{\alpha}\left(\tilde{V}_{nn}^{\alpha}\right)^{2}/\omega_{\alpha}^{2}$
is the reorganization energy (polaron shift) of the polaritonic state
$\left|\varphi_{n}\right\rangle $, $\lambda_{nm}=\frac{1}{2}\sum_{\alpha}W_{nm}^{\alpha}\left(\tilde{V}_{mm}^{\alpha}+\tilde{V}_{nn}^{\alpha}\right)/\omega_{\alpha}^{2}$,
and the rescaled couplings between the polaritonic states and the vibrational mode $\alpha$ is given by $W=e^{S}\tilde{V}e^{-S}$,
with matrix elements:
\begin{align}
W_{nm}^{\alpha} & =\exp\left(-\frac{i}{\hbar}\sum_{\gamma}\frac{p_{\gamma}\tilde{V}_{nn}^{\gamma}}{\omega_{\gamma}^{2}}\right)\nonumber \\
\times & \tilde{V}_{nm}^{\alpha}\exp\left(+\frac{i}{\hbar}\sum_{\gamma}\frac{p_{\gamma}\tilde{V}_{mm}^{\gamma}}{\omega_{\gamma}^{2}}\right).
\end{align}

The dynamics of the polaritonic excitation is approximated using the lowest order perturbation theory in the rescaled coupling term, $W_{nm}^{\alpha}$.
The equation of motion for the reduced density matrix representing the Hilbert space of the polaritonic states is given by the non--Markovian time--local secular Redfield equations. A detailed derivation and description of the non--Markovian equations is given in Supporting Information.

\begin{figure}[t]
\includegraphics[width=8cm]{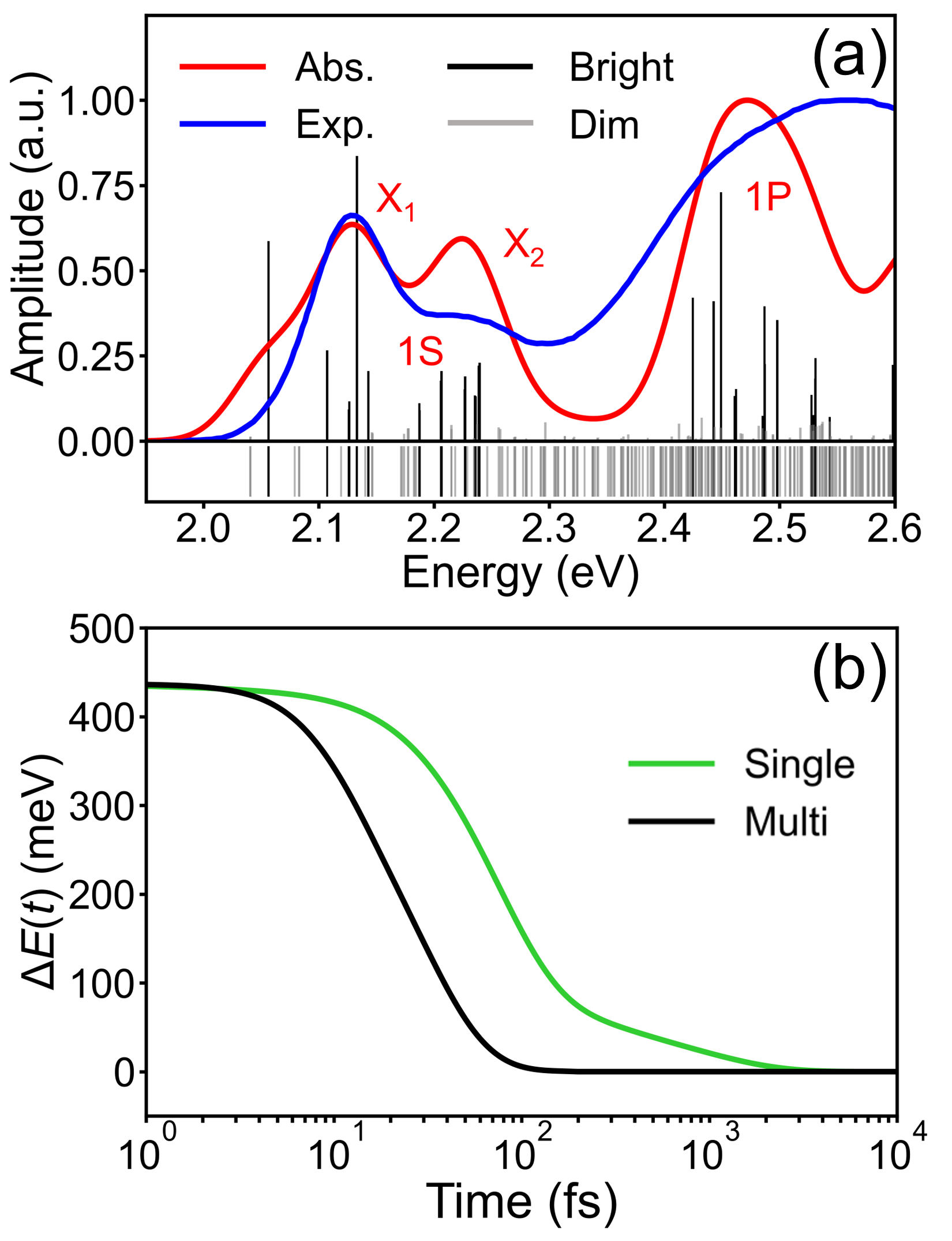}
\caption{(a) The calculated linear absorption spectrum (top) and the density of excitonic states (bottom) for a $\text{Cd}_{435}\text{Se}_{435}$ NC. The vertical lines in the top panel indicate the magnitude of the oscillator strength of the transition from the ground state to that excitonic state. (b) Phonon--mediated hot exciton 1P--1S cooling simulated for CdSe NCs. The inclusion of multiphonon processes allows the systems to relax to thermal equilibrium faster, circumventing
the phonon bottleneck.}

\label{fig:spectrum}
\end{figure}

\begin{figure*}[t]
\begin{centering}
\includegraphics[width=17.5cm]{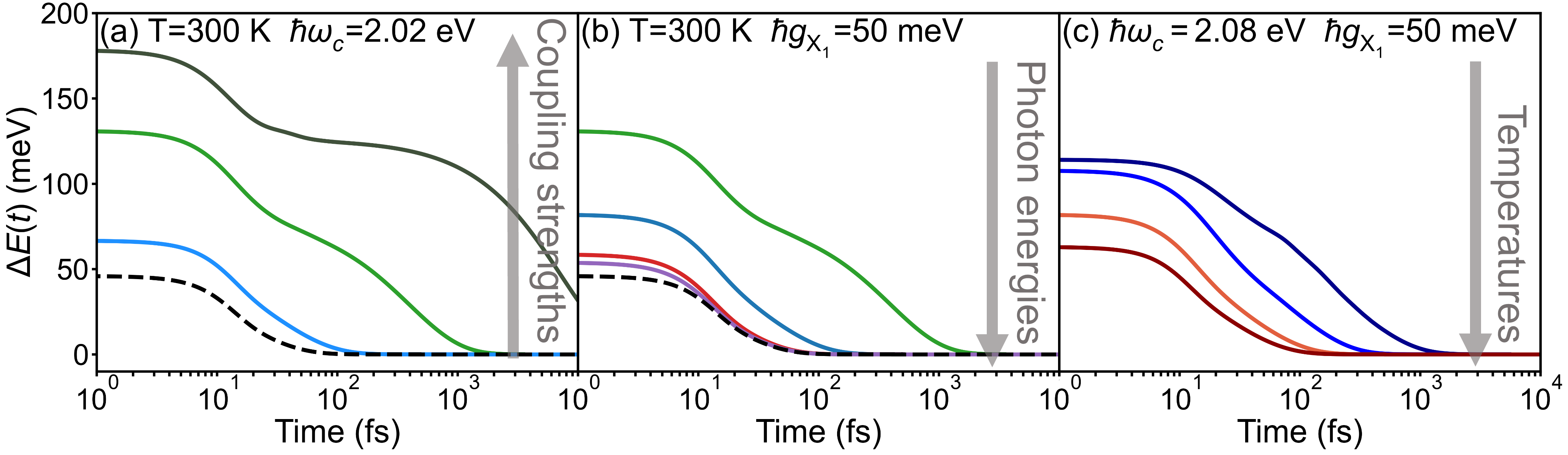}
\par\end{centering}
\centering{}\caption{Polaritonic energy relaxation for a $\text{Cd}_{435}\text{Se}_{435}$ NC inside a cavity. Panel (a) shows the dependence on the coupling strength for $\hbar g_{\text{X}_{1}}=0$ (black dashed), $\hbar g_{\text{X}_{1}}=20$ (light blue), $\hbar g_{\text{X}_{1}}=50$ (green), and $\hbar g_{\text{X}_{1}}=70$~meV (dark green) for $\hbar\omega_{c}=2.02$~eV (slightly below the absorption onset) and $T=300\text{\,K}$. Panel (b) shows the dependence on the photon energy for $\hbar\omega_{c}$ range from $2.02-2.20$~eV (in steps of $0.06$~eV), $\hbar g_{\text{X}_{1}}=50$~meV, and $T=300\,\text{K}$. Panel (c) shows the dependence on temperature for $T=50$K (dark blue), $150$K (blue), $300$K (orange), and $400$K (red), $\hbar\omega_{c}=2.08$~eV, and $\hbar g_{\text{X}_{1}}=50$~meV.} \label{fig:System_rela_all}
\end{figure*}

We consider a $\text{Cd}_{435}\text{Se}_{435}$ nanocrystal with a diameter of $D\approx4\thinspace\text{nm}$. The single particle states near the top of the valence band and the bottom of the conduction band were generated using the filter--diagonalization technique on a real--space grid basis with a total number of grid points exceeding $N_{g}>10^{6}$ with a grid spacing of $\approx0.65\thinspace a_{0}$. $80$ electron (unoccupied) and $140$ hole (occupied) states were calculated to construct the Bethe--Salpeter matrix. This was sufficient to converge the excitonic energies and transition dipole moments for the $\approx100$ lowest excitonic states, but only the lowest $\approx50$ were necessary to accurately describe the relaxation dynamics starting from the $\text{X}_{1}$ exciton (see Fig.~\ref{fig:spectrum}). In Fig.~\ref{fig:spectrum}(a) we plot the computed linear absorption spectrum of the $\text{Cd}_{435}\text{Se}_{435}$ nanocrystal alongside the density of excitonic states. The linear absorption spectrum shows several distinct features in agreement with the optical measurements (we label the main transitions as 1S and 1P following the literature convention).\cite{jasrasaria2022simulations} The density of excitonic states is relatively high due to the dense spectrum of hole states, but only a few excitonic states are "bright", as indicated by the oscillator strength of the individual transition also shown in Fig.~\ref{fig:spectrum}(a).
Most of the excitonic states are "dim" with a small oscillator strength.

The cooling dynamics of excitons for a $4\thinspace\text{nm}$ CdSe NC in the absence of a cavity mode was recently studied both theoretically and experimentally.\cite{jasrasaria2023circumventing,brosseau2023ultrafast} A mismatch between excitonic energy gaps and phonon frequencies has led to the hypothesis of a phonon bottleneck and extremely slow cooling, however, recent 2D electronic measurements revealed ultrafast excitonic relaxation on timescales of a few tens of femtoseconds, providing no evidence for a phonon bottleneck.\cite{brosseau2023ultrafast} The absence of a phonon bottleneck was rationalized by an Auger--assisted mechanism, which emerged naturally from the work of Jasrasaria and Rabani, however, it is the high density of vibrational modes and the coupling of excitons to both acoustic and optical modes that mainly contributes to the fast exciton cooling and the absence of a phonon bottleneck.\cite{jasrasaria2023circumventing}

\begin{figure*}[t!]
\includegraphics[width=17.5cm]{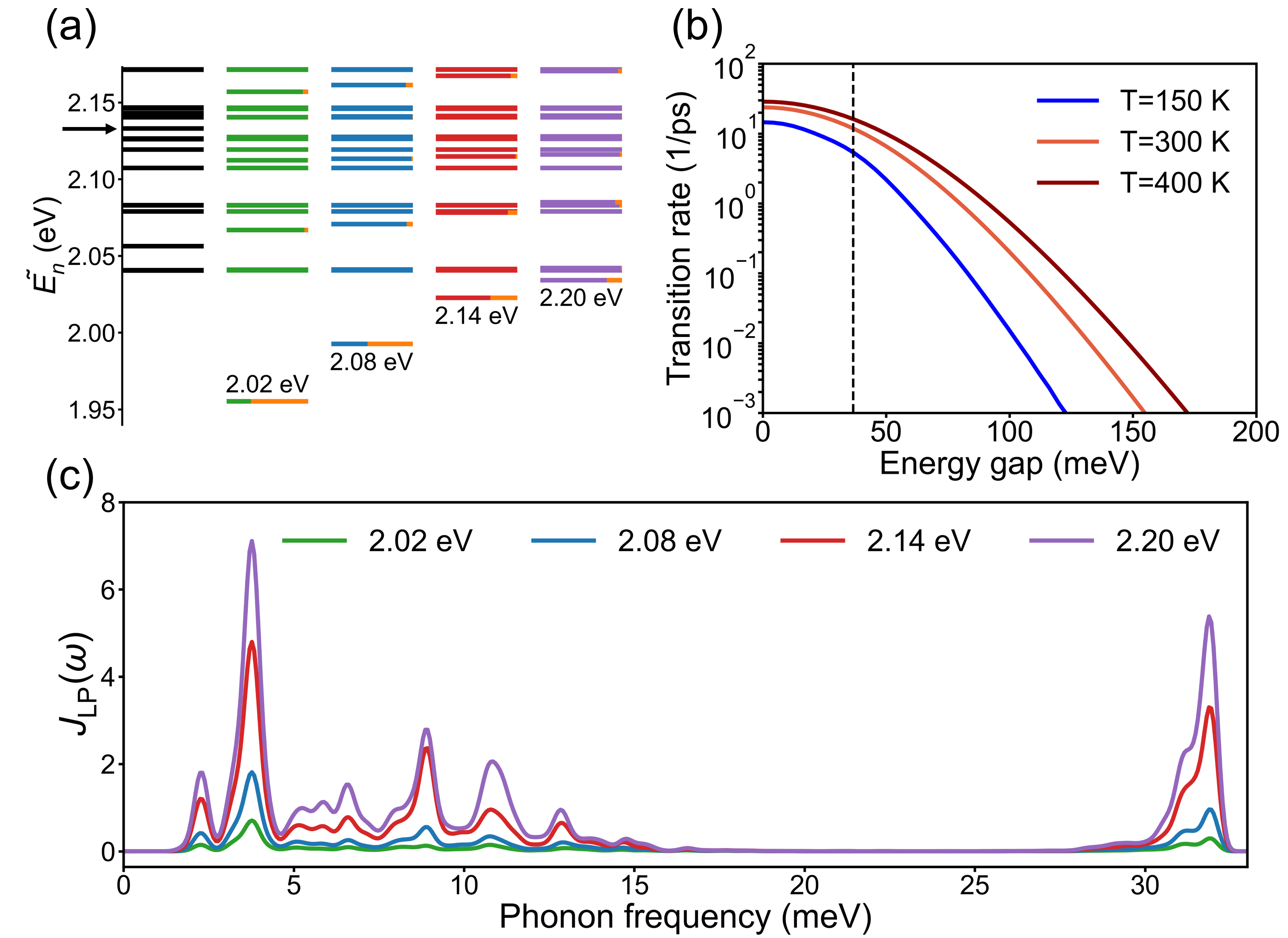}
\caption{(a) Polaritonic or excitonic energies for different cavity modes with a cavity--exciton coupling strength of $\hbar g_{\text{X}_{1}}=50$$\,$meV and $T=300\text{\,K}$. The photon fraction is depicted by the orange color for each photon energy ($\hbar\omega_{c}$ range from $2.02-2.20$~eV, in increments of $0.06$~eV).  The black arrow on the left side of the subplot is the initial excitonic state. (b) A semi--log plot of the transition rate between the two lowest polaritonic states as a function of the transition energy gap between them, for three different temperatures and for $\hbar\omega_{c}=2.08$~eV. The dashed curve shows this transition gap for $\text{Cd}_{435}\text{Se}_{435}$ at $\hbar\omega_{c}=2.08$~eV and $\hbar g_{\text{X}_{1}}=50$~meV. (c) The spectral density computed for the lowest polaritonic state as a function of phonon frequency for $4$ different photon energies. We use the same color code as in panel (a).}
\label{fig:DOS}
\end{figure*}

In Fig.~\ref{fig:spectrum}(b) we show the cooling dynamics when only single phonon processes are allowed (green curve) and for the multiphonon relaxation case (black curve). Following Ref.~\citenum{jasrasaria2023circumventing}, we show the relaxation across a wider energy range, starting from the 1P exciton. The single phonon processes were calculated using the same Redfield equation without performing the polaron transformation. We plot the average excitonic energy, $\Delta E\left(t\right)=\left\langle H_{\text{S}}\right\rangle \left(t\right)-\left\langle H_{\text{S}}\right\rangle _{{\rm eq}}$,\cite{satapathy2022thermalization} as a function of time, where $H_{\text{S}}$ is given in Eq.~\eqref{eq:hamil_S} for $g_{n}=0$. 

When only single phonon transitions are allowed, the 1P exciton relaxes on timescales that are much longer than observed experimentally.\cite{klimov1999electron,cooney2007unified} In this limit, the relaxation proceeds through a cascade of excitonic transitions with gaps that are smaller than the highest optical frequency. The slower long--time relaxation observed in this limit is a result of the sparse density of excitons near the band edge. For even smaller NCs ($D<3\thinspace\text{nm}$), the excitonic gaps become larger than the largest optical phonon frequency and the system cannot relax to thermal equilibrium in the limit of single--phonon processes.  

When multiphonon channels are allowed (black curve, Fig.~\ref{fig:spectrum}(b)), the decay occurs on much faster timescales and agrees well with recent measurements for both core and core/shell NCs.\cite{brosseau2023ultrafast} The inclusion of multiphonon processes allows for efficient relaxation of excitons across energy gaps that are larger than the largest optical frequency, opening multiphonon relaxation channels that compete with the single phonon paths, even for gaps that are smaller than the optical frequencies. The unexpected contribution of multiphonon channels to the relaxation of excitons was used to rationalize the absence of a phonon bottleneck.

In Fig.~\ref{fig:System_rela_all} we plot the relaxation dynamics for the same NC studied in Fig.~\ref{fig:spectrum} placed in an optical cavity starting from $\text{X}_{1}$.
We consider several different values of the coupling strengths (panel (a)), several different photon energies (panel (b)), and several different temperatures (panel (c)). 
For weak coupling to the cavity mode ($\hbar g_{\text{X}_{1}}<20$~meV), the relaxation timescales are similar to those computed for the isolated NC (dashed black curve). 
As the coupling to the cavity mode increases above $\hbar g_{\text{X}_{1}}>50$~meV, a slower timescale appears and the relaxation dynamics is characterized by a multi--exponential decay. 
A similar behavior was observed when either the photon energy
is significantly below the onset of the NC absorption (panel (b)) or at lower temperatures (panel (c)).
We note in passing that an analogous behavior was recently reported for a 2D quantum well system, where a significant slowing down of the relaxation dynamics was observed by manipulating the energy of the photon mode.\cite{laitz2023uncovering} 

To understand the emergence of this polaritonic--induced phonon bottleneck,
we plot in Fig.~\ref{fig:DOS}(a) the polaritonic energies ($\tilde{E_{n}}$) in the absence of a cavity (excitonic
energies, $E_{n}$) and for $4$ different photon energies. 
The cavity coupling is set to $\hbar g_{\text{X}_{1}}=50$~meV.
The lower polaritonic (LP) state has a significant photon fraction as depicted by the orange portion of the LP energy bar. 
Most of the polaritonic states above the LP have a vanishing fraction of the cavity mode (dim states couple very weakly to the cavity), except for a few bright states with non--vanishing photonic character.
The polaritonic density of states for different cavity--exciton coupling strengths is plotted in the Supporting Information (see Fig. S2(a)).

As the gap ($\delta E$) between the two lowest polaritonic states increases beyond the highest optical frequency, by either lowering the photon energy or increasing the coupling to the cavity, the relaxation to the lowest polaritonic state can only occur via multiphonon emission, with transition rates that are exponentially suppressed, as shown in Fig.~\ref{fig:DOS}(b). 
However, the increase in the gap between the two lowest polaritonic states with decreasing photon energy or increasing coupling, by itself, is not sufficient to account for the entire behavior shown in Fig.~\ref{fig:System_rela_all} panels (a) and (b). 

In Fig.~\ref{fig:DOS}(c) we plot the spectral density, $J_{\text{LP}}\left(\omega\right)=\frac{1}{2}\sum_{\alpha}\left(\frac{\tilde{V}_{\text{LP}}^{\alpha}}{\omega_{\alpha}}\right)^{2}\delta\left(\omega-\omega_{\alpha}\right)$, for LP for different values of the photon energy.  The spectral density is a measure of the strength of the polariton--phonon coupling to each phonon mode. As the cavity mode energy decreases below the absorption onset of the NC, the photon fraction of the LP state increases and the overall polariton--phonon coupling decreases. This, together with the opening of the gap $\delta E$ leads to a significant slowing down of the phonon emission rate and the emergence of a polaritonic--induced phonon bottleneck.

In the Supporting Information, we plot the spectral density for different values of the cavity--exciton coupling strengths (Fig.~S2(b)).  As the cavity--exciton coupling increases, the photon fraction of the LP state slightly decreases (as can be seen in Fig.~S2(a)), opposite to the behavior observed in Fig.~\ref{fig:DOS}(a) for different photon energies. As a result, the overall polariton--phonon coupling increases with increasing cavity--exciton couplings. This enhanced vibronic coupling competes with the slowing down of the relaxation due to the opening of the gap $\delta E$, but the dependence is rather mild to circumvent the bottleneck phenomena, as can be seen in population dynamics for different couplings strength shown in Fig.~\ref{fig:System_rela_all}(a).

To better quantify the onset of the polaritonic phonon bottleneck, we further analyze the transition rate between the two lowest polaritonic states, $\text{LP}$ and $\text{LP}+1$, as the bottleneck phenomenon is mainly correlated with the relaxation between these two states.
The decay rate between the two lowest polaritonic states with an energy spacing of $\delta E$ can be approximated by (see Supporting Information):
\begin{equation}
\varGamma\left(\delta E\right)\approx\frac{k_{{\rm B}}T}{\hbar}\sqrt{\frac{\pi\Lambda^{2}}{\eta\left(T\right)}}e^{-\delta E^{2}/4\eta\left(T\right)},\label{eq:exp_decay}
\end{equation}
where $\Lambda=\sum_{\alpha}\left|\frac{\tilde{V}^{\alpha}}{\omega_{\alpha}}\right|^{2}$
and $\eta\left(T\right)$ 
is a thermal weighted average measure of the polariton--phonon couplings: 

\begin{align}
\eta\left(T\right) & =-\frac{\hbar^{2}}{\Lambda}\sum_{\alpha}\left|\tilde{V}^{\alpha}\right|^{2}+\nonumber \\
+ & 2k_{{\rm B}}T\left(\lambda_{\text{LP}}+\lambda_{\text{LP+1}}-\sum_{\alpha}\frac{1}{\omega_{\alpha}^{2}}\tilde{V}_{\text{LP}}^{\alpha}\tilde{V}_{\text{LP+1}}^{\alpha}\right).
\end{align}

In the above equation, $\lambda_{\text{LP}}$ and $\lambda_{\text{LP+1}}$
are the reorganization energies of the lowest two polaritonic states, respectively,
$\tilde{V}_{\text{LP}}^{\alpha}$ and $\tilde{V}_{\text{LP+1}}^{\alpha}$
are the diagonal couplings between the lowest two polaritonic states and phonon mode $\alpha$, respectively, and $\tilde{V}^{\alpha}$ is the coupling between the two lowest polaritonic states to phonon
mode $\alpha$. 

\begin{figure}[t]
\includegraphics[width=8cm]{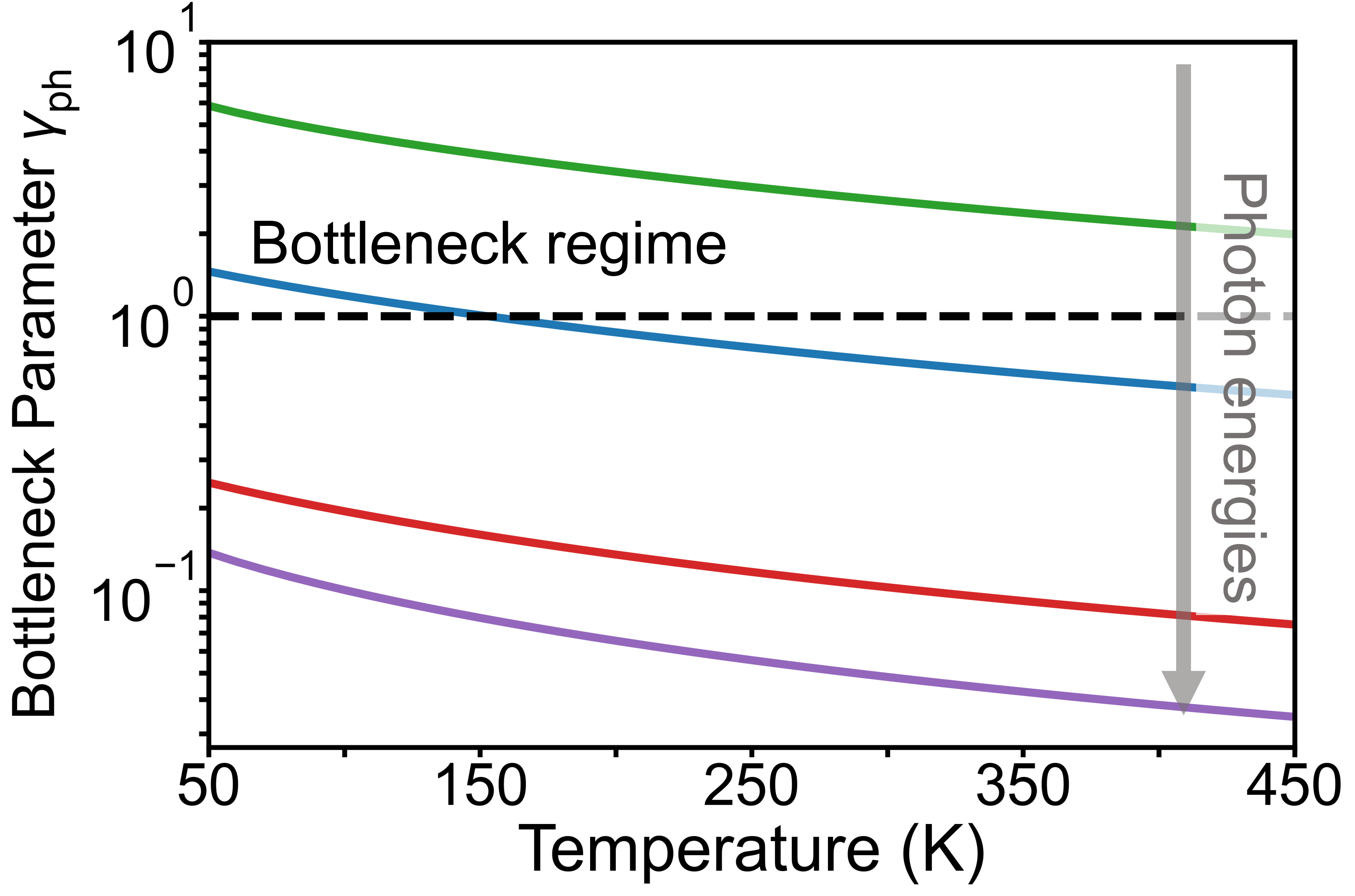}
\caption{Plot of $\gamma_{\rm ph} = \delta E^{2}/4\eta\left(T\right)$ vs. $T$ for $\hbar g_{\text{X}_{1}}=50$~meV and for $\hbar\omega_{c}$ range from $2.02-2.20$~eV, in increments
of $0.06$~eV. The dash black line is for $\gamma_{\rm ph} = 1$.
\label{fig:etaeta}}
\end{figure}

The dependence of transition rate on the $\delta E$ and $T$ is governed
by the exponential term ($e^{-\delta E^{2}/4\eta\left(T\right)}$),
while the pre--exponential factor ($\frac{k_{{\rm B}}T}{\hbar}\sqrt{\frac{\pi\Lambda^{2}}{\eta\left(T\right)}}$)
depends weakly on the optical cavity properties and temperature.  We define a "bottleneck parameter", $\gamma_{\rm ph} = \delta E^{2}/4\eta\left(T\right)$, and assume that the onset of a polaritonic phonon bottleneck occurs $\gamma_{\rm ph} > 1$ (dashed black curve in Fig.~\ref{fig:etaeta}). In Fig.~\ref{fig:etaeta} we plot $\gamma_{\rm ph}$ as a function of the temperature for $\hbar g_{\text{X}_{1}}=50$~meV and for all the values of the photon energy considered in this work. For different cavity--exciton coupling strengths, a similar diagram is shown in the Supporting Information, Fig. S4. The phonon bottleneck regime seems to appear for low cavity photon energies and for strong cavity--exciton coupling, while the temperatures can be used to drive the system in or out of a bottleneck regime.

In conclusion, we explored the relaxation of hot excitons in confined semiconductor NCs placed inside a micro--cavity. We adopted the pseudopotential model combined with the Bethe--Salpeter equation to describe the excitons and their coupling to the lattice phonons, and generalized the Pauli--Fierz Hamiltonian to account for the coupling of the excitons to a single cavity mode. Using the polaron--transformed Redfield equations within the polaritonic representation, we observe orders of magnitude inhibition of thermalization of polaritons by tuning the hybrid system, giving rise to a cavity--induced phonon bottleneck.  The transition to the bottleneck regime depends on the cavity photon mode, cavity--exciton coupling, and temperature. Our predictions provide fundamental insights into the nonradiative relaxation dynamics of confined excitons and how to control thermalization using optical cavities.

Supporting Information is available free of charge via the internet at http://pubs.acs.org.
Pauli--Fierz Hamiltonian, filter-diagonalization technique, Bethe-Salpeter equation, derivation of exciton--phonon coupling matrix elements, polaron transformation, non--Markovian local--time secular Redfield equation, additional discussion regarding polariton dynamics

\begin{acknowledgement}
This work was supported by the NSF-BSF International Collaboration in the Division of Materials Research program, NSF grant number DMR2026741. Computational resources were provided in part by the National Energy Research Scientific Computing Center (NERSC), a U.S. Department of Energy Office of Science User Facility operated under contract no. DEAC02- 05CH11231. We would like to thank Dr. Dipti Jasrasaria for helpful discussions. 
 
\end{acknowledgement}
\begin{tocentry}
\includegraphics[width=7cm]{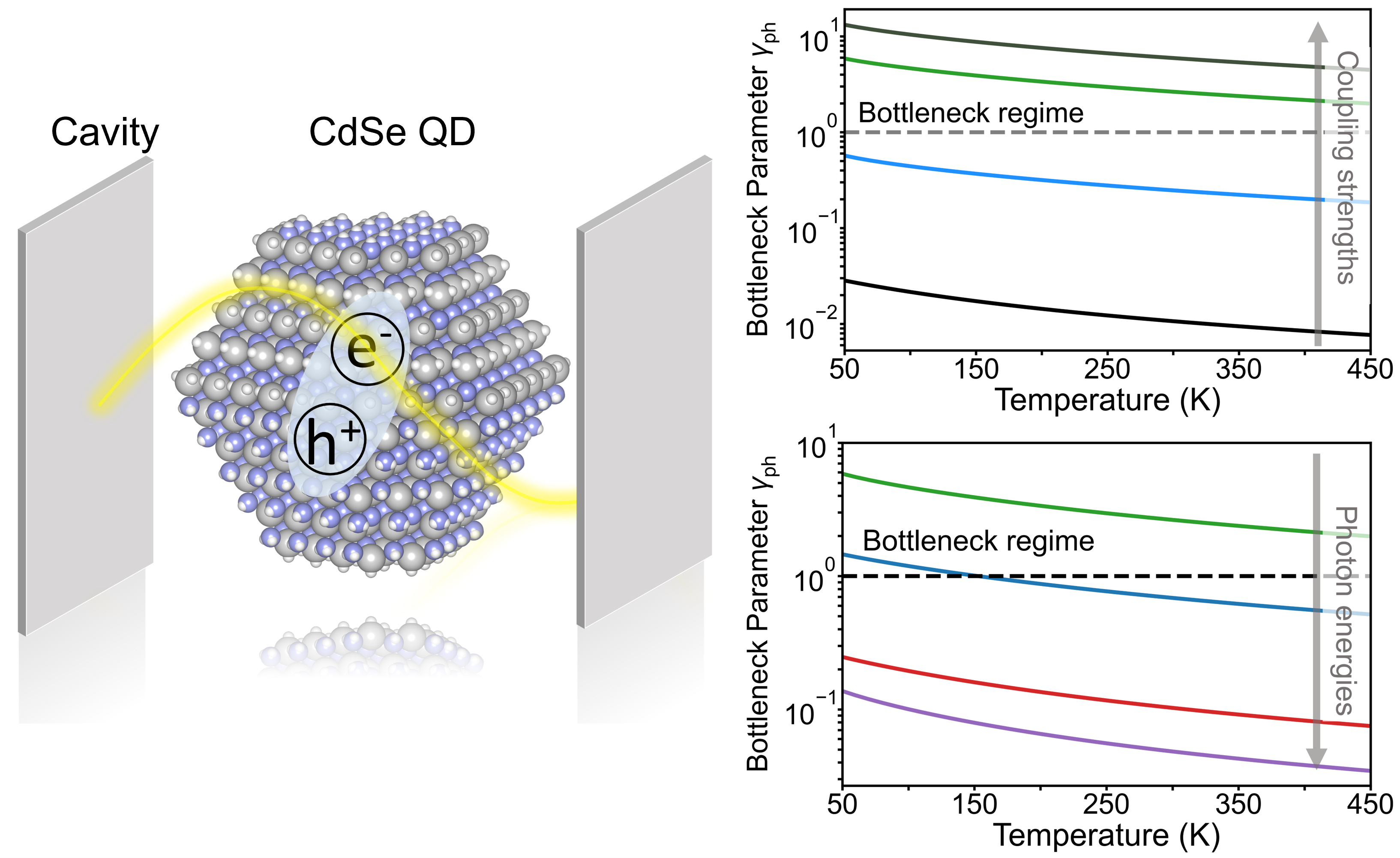}
\end{tocentry}
\bibliography{polariton}

\providecommand{\latin}[1]{#1}
\makeatletter
\providecommand{\doi}
  {\begingroup\let\do\@makeother\dospecials
  \catcode`\{=1 \catcode`\}=2 \doi@aux}
\providecommand{\doi@aux}[1]{\endgroup\texttt{#1}}
\makeatother
\providecommand*\mcitethebibliography{\thebibliography}
\csname @ifundefined\endcsname{endmcitethebibliography}
  {\let\endmcitethebibliography\endthebibliography}{}
\begin{mcitethebibliography}{47}
\providecommand*\natexlab[1]{#1}
\providecommand*\mciteSetBstSublistMode[1]{}
\providecommand*\mciteSetBstMaxWidthForm[2]{}
\providecommand*\mciteBstWouldAddEndPuncttrue
  {\def\EndOfBibitem{\unskip.}}
\providecommand*\mciteBstWouldAddEndPunctfalse
  {\let\EndOfBibitem\relax}
\providecommand*\mciteSetBstMidEndSepPunct[3]{}
\providecommand*\mciteSetBstSublistLabelBeginEnd[3]{}
\providecommand*\EndOfBibitem{}
\mciteSetBstSublistMode{f}
\mciteSetBstMaxWidthForm{subitem}{(\alph{mcitesubitemcount})}
\mciteSetBstSublistLabelBeginEnd
  {\mcitemaxwidthsubitemform\space}
  {\relax}
  {\relax}

\bibitem[Hou \latin{et~al.}(2023)Hou, Thoss, Banin, and
  Rabani]{hou_incoherent_2023}
Hou,~B.; Thoss,~M.; Banin,~U.; Rabani,~E. Incoherent nonadiabatic to coherent
  adiabatic transition of electron transfer in colloidal quantum dot molecules.
  \emph{Nat. Commun.} \textbf{2023}, \emph{14}, 3073\relax
\mciteBstWouldAddEndPuncttrue
\mciteSetBstMidEndSepPunct{\mcitedefaultmidpunct}
{\mcitedefaultendpunct}{\mcitedefaultseppunct}\relax
\EndOfBibitem
\bibitem[Bhattacharya \latin{et~al.}(2004)Bhattacharya, Ghosh, and
  Stiff-Roberts]{bhattacharya2004quantum}
Bhattacharya,~P.; Ghosh,~S.; Stiff-Roberts,~A. Quantum dot opto-electronic
  devices. \emph{Annu. Rev. Mater. Res.} \textbf{2004}, \emph{34}, 1--40\relax
\mciteBstWouldAddEndPuncttrue
\mciteSetBstMidEndSepPunct{\mcitedefaultmidpunct}
{\mcitedefaultendpunct}{\mcitedefaultseppunct}\relax
\EndOfBibitem
\bibitem[Hanifi \latin{et~al.}(2019)Hanifi, Bronstein, Koscher, Nett, Swabeck,
  Takano, Schwartzberg, Maserati, Vandewal, van~de Burgt, \latin{et~al.}
  others]{hanifi2019redefining}
Hanifi,~D.~A.; Bronstein,~N.~D.; Koscher,~B.~A.; Nett,~Z.; Swabeck,~J.~K.;
  Takano,~K.; Schwartzberg,~A.~M.; Maserati,~L.; Vandewal,~K.; van~de
  Burgt,~Y., \latin{et~al.}  Redefining near-unity luminescence in quantum dots
  with photothermal threshold quantum yield. \emph{Science} \textbf{2019},
  \emph{363}, 1199--1202\relax
\mciteBstWouldAddEndPuncttrue
\mciteSetBstMidEndSepPunct{\mcitedefaultmidpunct}
{\mcitedefaultendpunct}{\mcitedefaultseppunct}\relax
\EndOfBibitem
\bibitem[Philbin and Rabani(2020)Philbin, and Rabani]{philbin2020auger}
Philbin,~J.~P.; Rabani,~E. Auger recombination lifetime scaling for type I and
  quasi-type II core/shell quantum dots. \emph{J. Phys. Chem. Lett.}
  \textbf{2020}, \emph{11}, 5132--5138\relax
\mciteBstWouldAddEndPuncttrue
\mciteSetBstMidEndSepPunct{\mcitedefaultmidpunct}
{\mcitedefaultendpunct}{\mcitedefaultseppunct}\relax
\EndOfBibitem
\bibitem[Jasrasaria \latin{et~al.}(2022)Jasrasaria, Weinberg, Philbin, and
  Rabani]{jasrasaria2022simulations}
Jasrasaria,~D.; Weinberg,~D.; Philbin,~J.~P.; Rabani,~E. Simulations of
  nonradiative processes in semiconductor nanocrystals. \emph{J. Chem. Phys.}
  \textbf{2022}, \emph{157}, 020901\relax
\mciteBstWouldAddEndPuncttrue
\mciteSetBstMidEndSepPunct{\mcitedefaultmidpunct}
{\mcitedefaultendpunct}{\mcitedefaultseppunct}\relax
\EndOfBibitem
\bibitem[Melnychuk and Guyot-Sionnest(2021)Melnychuk, and
  Guyot-Sionnest]{melnychuk2021multicarrier}
Melnychuk,~C.; Guyot-Sionnest,~P. Multicarrier dynamics in quantum dots.
  \emph{Chem. Rev.} \textbf{2021}, \emph{121}, 2325--2372\relax
\mciteBstWouldAddEndPuncttrue
\mciteSetBstMidEndSepPunct{\mcitedefaultmidpunct}
{\mcitedefaultendpunct}{\mcitedefaultseppunct}\relax
\EndOfBibitem
\bibitem[Schaller \latin{et~al.}(2005)Schaller, Pietryga, Goupalov, Petruska,
  Ivanov, and Klimov]{schaller2005breaking}
Schaller,~R.~D.; Pietryga,~J.~M.; Goupalov,~S.~V.; Petruska,~M.~A.;
  Ivanov,~S.~A.; Klimov,~V.~I. Breaking the phonon bottleneck in semiconductor
  nanocrystals via multiphonon emission induced by intrinsic nonadiabatic
  interactions. \emph{Phys. Rev. Lett.} \textbf{2005}, \emph{95}, 196401\relax
\mciteBstWouldAddEndPuncttrue
\mciteSetBstMidEndSepPunct{\mcitedefaultmidpunct}
{\mcitedefaultendpunct}{\mcitedefaultseppunct}\relax
\EndOfBibitem
\bibitem[Cooney \latin{et~al.}(2007)Cooney, Sewall, Anderson, Dias, and
  Kambhampati]{cooney2007breaking}
Cooney,~R.~R.; Sewall,~S.~L.; Anderson,~K.~E.; Dias,~E.~A.; Kambhampati,~P.
  Breaking the phonon bottleneck for holes in semiconductor quantum dots.
  \emph{Phys. Rev. Lett.} \textbf{2007}, \emph{98}, 177403\relax
\mciteBstWouldAddEndPuncttrue
\mciteSetBstMidEndSepPunct{\mcitedefaultmidpunct}
{\mcitedefaultendpunct}{\mcitedefaultseppunct}\relax
\EndOfBibitem
\bibitem[Cooney \latin{et~al.}(2007)Cooney, Sewall, Dias, Sagar, Anderson, and
  Kambhampati]{cooney2007unified}
Cooney,~R.~R.; Sewall,~S.~L.; Dias,~E.~A.; Sagar,~D.; Anderson,~K.~E.;
  Kambhampati,~P. Unified picture of electron and hole relaxation pathways in
  semiconductor quantum dots. \emph{Phys. Rev. B} \textbf{2007}, \emph{75},
  245311\relax
\mciteBstWouldAddEndPuncttrue
\mciteSetBstMidEndSepPunct{\mcitedefaultmidpunct}
{\mcitedefaultendpunct}{\mcitedefaultseppunct}\relax
\EndOfBibitem
\bibitem[Jasrasaria and Rabani(2023)Jasrasaria, and
  Rabani]{jasrasaria2023circumventing}
Jasrasaria,~D.; Rabani,~E. Circumventing the phonon bottleneck by
  multiphonon-mediated hot exciton cooling at the nanoscale. \emph{Npj Comput.
  Mater.} \textbf{2023}, \emph{9}, 1--8\relax
\mciteBstWouldAddEndPuncttrue
\mciteSetBstMidEndSepPunct{\mcitedefaultmidpunct}
{\mcitedefaultendpunct}{\mcitedefaultseppunct}\relax
\EndOfBibitem
\bibitem[Hu \latin{et~al.}(2022)Hu, Gustin, Krauss, and Franco]{hu2022tuning}
Hu,~W.; Gustin,~I.; Krauss,~T.~D.; Franco,~I. Tuning and Enhancing Quantum
  Coherence Time Scales in Molecules via Light-Matter Hybridization. \emph{J.
  Phys. Chem. Lett.} \textbf{2022}, \emph{13}, 11503--11511\relax
\mciteBstWouldAddEndPuncttrue
\mciteSetBstMidEndSepPunct{\mcitedefaultmidpunct}
{\mcitedefaultendpunct}{\mcitedefaultseppunct}\relax
\EndOfBibitem
\bibitem[Fomenko and Nesbitt(2008)Fomenko, and Nesbitt]{fomenko2008solution}
Fomenko,~V.; Nesbitt,~D.~J. Solution control of radiative and nonradiative
  lifetimes: A novel contribution to quantum dot blinking suppression.
  \emph{Nano Lett.} \textbf{2008}, \emph{8}, 287--293\relax
\mciteBstWouldAddEndPuncttrue
\mciteSetBstMidEndSepPunct{\mcitedefaultmidpunct}
{\mcitedefaultendpunct}{\mcitedefaultseppunct}\relax
\EndOfBibitem
\bibitem[Talapin \latin{et~al.}(2010)Talapin, Lee, Kovalenko, and
  Shevchenko]{talapin2010prospects}
Talapin,~D.~V.; Lee,~J.-S.; Kovalenko,~M.~V.; Shevchenko,~E.~V. Prospects of
  colloidal nanocrystals for electronic and optoelectronic applications.
  \emph{Chem. Rev.} \textbf{2010}, \emph{110}, 389--458\relax
\mciteBstWouldAddEndPuncttrue
\mciteSetBstMidEndSepPunct{\mcitedefaultmidpunct}
{\mcitedefaultendpunct}{\mcitedefaultseppunct}\relax
\EndOfBibitem
\bibitem[Jasrasaria and Rabani(2021)Jasrasaria, and
  Rabani]{jasrasaria2021interplay}
Jasrasaria,~D.; Rabani,~E. Interplay of surface and interior modes in
  exciton--phonon coupling at the nanoscale. \emph{Nano Lett.} \textbf{2021},
  \emph{21}, 8741--8748\relax
\mciteBstWouldAddEndPuncttrue
\mciteSetBstMidEndSepPunct{\mcitedefaultmidpunct}
{\mcitedefaultendpunct}{\mcitedefaultseppunct}\relax
\EndOfBibitem
\bibitem[Sch{\"a}fer \latin{et~al.}(2019)Sch{\"a}fer, Ruggenthaler, Appel, and
  Rubio]{schafer2019modification}
Sch{\"a}fer,~C.; Ruggenthaler,~M.; Appel,~H.; Rubio,~A. Modification of
  excitation and charge transfer in cavity quantum-electrodynamical chemistry.
  \emph{Proc. Natl. Acad. Sci.} \textbf{2019}, \emph{116}, 4883--4892\relax
\mciteBstWouldAddEndPuncttrue
\mciteSetBstMidEndSepPunct{\mcitedefaultmidpunct}
{\mcitedefaultendpunct}{\mcitedefaultseppunct}\relax
\EndOfBibitem
\bibitem[Du \latin{et~al.}(2018)Du, Mart{\'\i}nez-Mart{\'\i}nez, Ribeiro, Hu,
  Menon, and Yuen-Zhou]{du2018theory}
Du,~M.; Mart{\'\i}nez-Mart{\'\i}nez,~L.~A.; Ribeiro,~R.~F.; Hu,~Z.;
  Menon,~V.~M.; Yuen-Zhou,~J. Theory for polariton-assisted remote energy
  transfer. \emph{Chem. Sci.} \textbf{2018}, \emph{9}, 6659--6669\relax
\mciteBstWouldAddEndPuncttrue
\mciteSetBstMidEndSepPunct{\mcitedefaultmidpunct}
{\mcitedefaultendpunct}{\mcitedefaultseppunct}\relax
\EndOfBibitem
\bibitem[Gonzalez-Ballestero \latin{et~al.}(2015)Gonzalez-Ballestero, Feist,
  Moreno, and Garcia-Vidal]{gonzalez2015harvesting}
Gonzalez-Ballestero,~C.; Feist,~J.; Moreno,~E.; Garcia-Vidal,~F.~J. Harvesting
  excitons through plasmonic strong coupling. \emph{Phys. Rev. B}
  \textbf{2015}, \emph{92}, 121402\relax
\mciteBstWouldAddEndPuncttrue
\mciteSetBstMidEndSepPunct{\mcitedefaultmidpunct}
{\mcitedefaultendpunct}{\mcitedefaultseppunct}\relax
\EndOfBibitem
\bibitem[Westmoreland \latin{et~al.}(2019)Westmoreland, McClelland, Perez,
  Schwabacher, Zhang, and Weiss]{westmoreland2019properties}
Westmoreland,~D.~E.; McClelland,~K.~P.; Perez,~K.~A.; Schwabacher,~J.~C.;
  Zhang,~Z.; Weiss,~E.~A. Properties of quantum dots coupled to plasmons and
  optical cavities. \emph{J. Chem. Phys.} \textbf{2019}, \emph{151},
  210901\relax
\mciteBstWouldAddEndPuncttrue
\mciteSetBstMidEndSepPunct{\mcitedefaultmidpunct}
{\mcitedefaultendpunct}{\mcitedefaultseppunct}\relax
\EndOfBibitem
\bibitem[Coles \latin{et~al.}(2013)Coles, Grant, Lidzey, Clark, and
  Lagoudakis]{coles2013imaging}
Coles,~D.~M.; Grant,~R.~T.; Lidzey,~D.~G.; Clark,~C.; Lagoudakis,~P.~G. Imaging
  the polariton relaxation bottleneck in strongly coupled organic semiconductor
  microcavities. \emph{Phys. Rev. B} \textbf{2013}, \emph{88}, 121303\relax
\mciteBstWouldAddEndPuncttrue
\mciteSetBstMidEndSepPunct{\mcitedefaultmidpunct}
{\mcitedefaultendpunct}{\mcitedefaultseppunct}\relax
\EndOfBibitem
\bibitem[Laitz \latin{et~al.}(2023)Laitz, Kaplan, Deschamps, Barotov, Proppe,
  Garc{\'\i}a-Benito, Osherov, Grancini, deQuilettes, Nelson, \latin{et~al.}
  others]{laitz2023uncovering}
Laitz,~M.; Kaplan,~A.~E.; Deschamps,~J.; Barotov,~U.; Proppe,~A.~H.;
  Garc{\'\i}a-Benito,~I.; Osherov,~A.; Grancini,~G.; deQuilettes,~D.~W.;
  Nelson,~K.~A., \latin{et~al.}  Uncovering temperature-dependent
  exciton-polariton relaxation mechanisms in hybrid organic-inorganic
  perovskites. \emph{Nat. Commun.} \textbf{2023}, \emph{14}, 2426\relax
\mciteBstWouldAddEndPuncttrue
\mciteSetBstMidEndSepPunct{\mcitedefaultmidpunct}
{\mcitedefaultendpunct}{\mcitedefaultseppunct}\relax
\EndOfBibitem
\bibitem[Michetti and La~Rocca(2008)Michetti, and
  La~Rocca]{michetti2008simulation}
Michetti,~P.; La~Rocca,~G.~C. Simulation of J-aggregate microcavity
  photoluminescence. \emph{Phys. Rev. B} \textbf{2008}, \emph{77}, 195301\relax
\mciteBstWouldAddEndPuncttrue
\mciteSetBstMidEndSepPunct{\mcitedefaultmidpunct}
{\mcitedefaultendpunct}{\mcitedefaultseppunct}\relax
\EndOfBibitem
\bibitem[Stokker-Cheregi \latin{et~al.}(2008)Stokker-Cheregi, Vinattieri,
  Semond, Leroux, Sellers, Massies, Solnyshkov, Malpuech, Colocci, and
  Gurioli]{stokker2008polariton}
Stokker-Cheregi,~F.; Vinattieri,~A.; Semond,~F.; Leroux,~M.; Sellers,~I.;
  Massies,~J.; Solnyshkov,~D.; Malpuech,~G.; Colocci,~M.; Gurioli,~M. Polariton
  relaxation bottleneck and its thermal suppression in bulk GaN microcavities.
  \emph{Appl. Phys. Lett.} \textbf{2008}, \emph{92}, 042119\relax
\mciteBstWouldAddEndPuncttrue
\mciteSetBstMidEndSepPunct{\mcitedefaultmidpunct}
{\mcitedefaultendpunct}{\mcitedefaultseppunct}\relax
\EndOfBibitem
\bibitem[Mandal and Huo(2019)Mandal, and Huo]{mandal2019investigating}
Mandal,~A.; Huo,~P. Investigating new reactivities enabled by polariton
  photochemistry. \emph{J. Phys. Chem. Lett.} \textbf{2019}, \emph{10},
  5519--5529\relax
\mciteBstWouldAddEndPuncttrue
\mciteSetBstMidEndSepPunct{\mcitedefaultmidpunct}
{\mcitedefaultendpunct}{\mcitedefaultseppunct}\relax
\EndOfBibitem
\bibitem[Sch{\"a}fer \latin{et~al.}(2022)Sch{\"a}fer, Flick, Ronca, Narang, and
  Rubio]{schafer2022shining}
Sch{\"a}fer,~C.; Flick,~J.; Ronca,~E.; Narang,~P.; Rubio,~A. Shining light on
  the microscopic resonant mechanism responsible for cavity-mediated chemical
  reactivity. \emph{Nat. Commun.} \textbf{2022}, \emph{13}, 7817\relax
\mciteBstWouldAddEndPuncttrue
\mciteSetBstMidEndSepPunct{\mcitedefaultmidpunct}
{\mcitedefaultendpunct}{\mcitedefaultseppunct}\relax
\EndOfBibitem
\bibitem[K{\'e}na-Cohen and Forrest(2010)K{\'e}na-Cohen, and
  Forrest]{kena2010room}
K{\'e}na-Cohen,~S.; Forrest,~S. Room-temperature polariton lasing in an organic
  single-crystal microcavity. \emph{Nat. Photonics} \textbf{2010}, \emph{4},
  371--375\relax
\mciteBstWouldAddEndPuncttrue
\mciteSetBstMidEndSepPunct{\mcitedefaultmidpunct}
{\mcitedefaultendpunct}{\mcitedefaultseppunct}\relax
\EndOfBibitem
\bibitem[Mazza \latin{et~al.}(2013)Mazza, K{\'e}na-Cohen, Michetti, and
  La~Rocca]{mazza2013microscopic}
Mazza,~L.; K{\'e}na-Cohen,~S.; Michetti,~P.; La~Rocca,~G.~C. Microscopic theory
  of polariton lasing via vibronically assisted scattering. \emph{Phys. Rev. B}
  \textbf{2013}, \emph{88}, 075321\relax
\mciteBstWouldAddEndPuncttrue
\mciteSetBstMidEndSepPunct{\mcitedefaultmidpunct}
{\mcitedefaultendpunct}{\mcitedefaultseppunct}\relax
\EndOfBibitem
\bibitem[Fraser \latin{et~al.}(2016)Fraser, H{\"o}fling, and
  Yamamoto]{fraser2016physics}
Fraser,~M.~D.; H{\"o}fling,~S.; Yamamoto,~Y. Physics and applications of
  exciton--polariton lasers. \emph{Nat. Mater.} \textbf{2016}, \emph{15},
  1049--1052\relax
\mciteBstWouldAddEndPuncttrue
\mciteSetBstMidEndSepPunct{\mcitedefaultmidpunct}
{\mcitedefaultendpunct}{\mcitedefaultseppunct}\relax
\EndOfBibitem
\bibitem[Imamog \latin{et~al.}(1996)Imamog, Ram, Pau, Yamamoto, \latin{et~al.}
  others]{imamog1996nonequilibrium}
Imamog,~A.; Ram,~R.; Pau,~S.; Yamamoto,~Y., \latin{et~al.}  Nonequilibrium
  condensates and lasers without inversion: Exciton-polariton lasers.
  \emph{Phys. Rev. A} \textbf{1996}, \emph{53}, 4250\relax
\mciteBstWouldAddEndPuncttrue
\mciteSetBstMidEndSepPunct{\mcitedefaultmidpunct}
{\mcitedefaultendpunct}{\mcitedefaultseppunct}\relax
\EndOfBibitem
\bibitem[Bajoni(2012)]{bajoni2012polariton}
Bajoni,~D. Polariton lasers. Hybrid light--matter lasers without inversion.
  \emph{J. Phys. D: Appl. Phys.} \textbf{2012}, \emph{45}, 313001\relax
\mciteBstWouldAddEndPuncttrue
\mciteSetBstMidEndSepPunct{\mcitedefaultmidpunct}
{\mcitedefaultendpunct}{\mcitedefaultseppunct}\relax
\EndOfBibitem
\bibitem[Amo \latin{et~al.}(2009)Amo, Sanvitto, Laussy, Ballarini, Valle,
  Martin, Lemaitre, Bloch, Krizhanovskii, Skolnick, \latin{et~al.}
  others]{amo2009collective}
Amo,~A.; Sanvitto,~D.; Laussy,~F.; Ballarini,~D.; Valle,~E.~d.; Martin,~M.;
  Lemaitre,~A.; Bloch,~J.; Krizhanovskii,~D.; Skolnick,~M., \latin{et~al.}
  Collective fluid dynamics of a polariton condensate in a semiconductor
  microcavity. \emph{Nature} \textbf{2009}, \emph{457}, 291--295\relax
\mciteBstWouldAddEndPuncttrue
\mciteSetBstMidEndSepPunct{\mcitedefaultmidpunct}
{\mcitedefaultendpunct}{\mcitedefaultseppunct}\relax
\EndOfBibitem
\bibitem[Kasprzak \latin{et~al.}(2006)Kasprzak, Richard, Kundermann, Baas,
  Jeambrun, Keeling, Marchetti, Szyma{\'n}ska, Andr{\'e}, Staehli,
  \latin{et~al.} others]{kasprzak2006bose}
Kasprzak,~J.; Richard,~M.; Kundermann,~S.; Baas,~A.; Jeambrun,~P.; Keeling,~J.
  M.~J.; Marchetti,~F.; Szyma{\'n}ska,~M.; Andr{\'e},~R.; Staehli,~J.,
  \latin{et~al.}  Bose--Einstein condensation of exciton polaritons.
  \emph{Nature} \textbf{2006}, \emph{443}, 409--414\relax
\mciteBstWouldAddEndPuncttrue
\mciteSetBstMidEndSepPunct{\mcitedefaultmidpunct}
{\mcitedefaultendpunct}{\mcitedefaultseppunct}\relax
\EndOfBibitem
\bibitem[Deng \latin{et~al.}(2010)Deng, Haug, and Yamamoto]{deng2010exciton}
Deng,~H.; Haug,~H.; Yamamoto,~Y. Exciton-polariton bose-einstein condensation.
  \emph{Rev. Mod. Phys.} \textbf{2010}, \emph{82}, 1489\relax
\mciteBstWouldAddEndPuncttrue
\mciteSetBstMidEndSepPunct{\mcitedefaultmidpunct}
{\mcitedefaultendpunct}{\mcitedefaultseppunct}\relax
\EndOfBibitem
\bibitem[Ghosh and Liew(2020)Ghosh, and Liew]{ghosh2020quantum}
Ghosh,~S.; Liew,~T.~C. Quantum computing with exciton-polariton condensates.
  \emph{Npj Quantum Inf.} \textbf{2020}, \emph{6}, 16\relax
\mciteBstWouldAddEndPuncttrue
\mciteSetBstMidEndSepPunct{\mcitedefaultmidpunct}
{\mcitedefaultendpunct}{\mcitedefaultseppunct}\relax
\EndOfBibitem
\bibitem[Xue \latin{et~al.}(2021)Xue, Chestnov, Sedov, Kiktenko, Fedorov,
  Schumacher, Ma, and Kavokin]{xue2021split}
Xue,~Y.; Chestnov,~I.; Sedov,~E.; Kiktenko,~E.; Fedorov,~A.~K.; Schumacher,~S.;
  Ma,~X.; Kavokin,~A. Split-ring polariton condensates as macroscopic two-level
  quantum systems. \emph{Phys. Rev. Res.} \textbf{2021}, \emph{3}, 013099\relax
\mciteBstWouldAddEndPuncttrue
\mciteSetBstMidEndSepPunct{\mcitedefaultmidpunct}
{\mcitedefaultendpunct}{\mcitedefaultseppunct}\relax
\EndOfBibitem
\bibitem[Demirchyan \latin{et~al.}(2014)Demirchyan, Chestnov, Alodjants,
  Glazov, and Kavokin]{demirchyan2014qubits}
Demirchyan,~S.; Chestnov,~I.~Y.; Alodjants,~A.; Glazov,~M.; Kavokin,~A. Qubits
  based on polariton Rabi oscillators. \emph{Phys. Rev. Lett.} \textbf{2014},
  \emph{112}, 196403\relax
\mciteBstWouldAddEndPuncttrue
\mciteSetBstMidEndSepPunct{\mcitedefaultmidpunct}
{\mcitedefaultendpunct}{\mcitedefaultseppunct}\relax
\EndOfBibitem
\bibitem[Mandal \latin{et~al.}(2023)Mandal, Taylor, Weight, Koessler, Li, and
  Huo]{mandal_theoretical_2023}
Mandal,~A.; Taylor,~M.~A.; Weight,~B.~M.; Koessler,~E.~R.; Li,~X.; Huo,~P.
  Theoretical Advances in Polariton Chemistry and Molecular Cavity Quantum
  Electrodynamics. \emph{Chem. Rev.} \textbf{2023}, \emph{123},
  9786--9879\relax
\mciteBstWouldAddEndPuncttrue
\mciteSetBstMidEndSepPunct{\mcitedefaultmidpunct}
{\mcitedefaultendpunct}{\mcitedefaultseppunct}\relax
\EndOfBibitem
\bibitem[Xu and Cao(2016)Xu, and Cao]{xu2016non}
Xu,~D.; Cao,~J. Non-canonical distribution and non-equilibrium transport beyond
  weak system-bath coupling regime: A polaron transformation approach.
  \emph{Front. Phys-beijing.} \textbf{2016}, \emph{11}, 1--17\relax
\mciteBstWouldAddEndPuncttrue
\mciteSetBstMidEndSepPunct{\mcitedefaultmidpunct}
{\mcitedefaultendpunct}{\mcitedefaultseppunct}\relax
\EndOfBibitem
\bibitem[Zhou \latin{et~al.}(2013)Zhou, Ward, Martin, Van~Swol, Cruz-Campa, and
  Zubia]{zhou2013stillinger}
Zhou,~X.; Ward,~D.; Martin,~J.; Van~Swol,~F.; Cruz-Campa,~J.; Zubia,~D.
  Stillinger-weber potential for the II-VI elements zn-cd-hg-S-se-te.
  \emph{Phys. Rev. B} \textbf{2013}, \emph{88}, 085309\relax
\mciteBstWouldAddEndPuncttrue
\mciteSetBstMidEndSepPunct{\mcitedefaultmidpunct}
{\mcitedefaultendpunct}{\mcitedefaultseppunct}\relax
\EndOfBibitem
\bibitem[Rabani \latin{et~al.}(1999)Rabani, Hetenyi, Berne, and
  Brus]{rabani1999electronic}
Rabani,~E.; Hetenyi,~B.; Berne,~B.~J.; Brus,~L.~E. Electronic properties of
  CdSe nanocrystals in the absence and presence of a dielectric medium.
  \emph{J. Chem. Phys.} \textbf{1999}, \emph{110}, 5355--5369\relax
\mciteBstWouldAddEndPuncttrue
\mciteSetBstMidEndSepPunct{\mcitedefaultmidpunct}
{\mcitedefaultendpunct}{\mcitedefaultseppunct}\relax
\EndOfBibitem
\bibitem[Rohlfing and Louie(2000)Rohlfing, and Louie]{rohlfing2000electron}
Rohlfing,~M.; Louie,~S.~G. Electron-hole excitations and optical spectra from
  first principles. \emph{Phys. Rev. B} \textbf{2000}, \emph{62}, 4927\relax
\mciteBstWouldAddEndPuncttrue
\mciteSetBstMidEndSepPunct{\mcitedefaultmidpunct}
{\mcitedefaultendpunct}{\mcitedefaultseppunct}\relax
\EndOfBibitem
\bibitem[Lin \latin{et~al.}(2023)Lin, Jasrasaria, Yoo, Bawendi, Utzat, and
  Rabani]{doi:10.1021/acs.jpclett.3c01630}
Lin,~K.; Jasrasaria,~D.; Yoo,~J.~J.; Bawendi,~M.; Utzat,~H.; Rabani,~E. Theory
  of Photoluminescence Spectral Line Shapes of Semiconductor Nanocrystals.
  \emph{J. Phys. Chem. Lett.} \textbf{2023}, \emph{14}, 7241--7248\relax
\mciteBstWouldAddEndPuncttrue
\mciteSetBstMidEndSepPunct{\mcitedefaultmidpunct}
{\mcitedefaultendpunct}{\mcitedefaultseppunct}\relax
\EndOfBibitem
\bibitem[Brosseau \latin{et~al.}(2023)Brosseau, Geuchies, Jasrasaria, Houtepen,
  Rabani, and Kambhampati]{brosseau2023ultrafast}
Brosseau,~P.~J.; Geuchies,~J.~J.; Jasrasaria,~D.; Houtepen,~A.~J.; Rabani,~E.;
  Kambhampati,~P. Ultrafast hole relaxation dynamics in quantum dots revealed
  by two-dimensional electronic spectroscopy. \emph{Commun. Phys.}
  \textbf{2023}, \emph{6}, 48\relax
\mciteBstWouldAddEndPuncttrue
\mciteSetBstMidEndSepPunct{\mcitedefaultmidpunct}
{\mcitedefaultendpunct}{\mcitedefaultseppunct}\relax
\EndOfBibitem
\bibitem[Mandal \latin{et~al.}(2020)Mandal, Krauss, and
  Huo]{mandal2020polariton}
Mandal,~A.; Krauss,~T.~D.; Huo,~P. Polariton-mediated electron transfer via
  cavity quantum electrodynamics. \emph{J. Phys. Chem. B} \textbf{2020},
  \emph{124}, 6321--6340\relax
\mciteBstWouldAddEndPuncttrue
\mciteSetBstMidEndSepPunct{\mcitedefaultmidpunct}
{\mcitedefaultendpunct}{\mcitedefaultseppunct}\relax
\EndOfBibitem
\bibitem[Philbin and Rabani(2018)Philbin, and Rabani]{philbin2018electron}
Philbin,~J.~P.; Rabani,~E. Electron--hole correlations govern auger
  recombination in nanostructures. \emph{Nano Lett.} \textbf{2018}, \emph{18},
  7889--7895\relax
\mciteBstWouldAddEndPuncttrue
\mciteSetBstMidEndSepPunct{\mcitedefaultmidpunct}
{\mcitedefaultendpunct}{\mcitedefaultseppunct}\relax
\EndOfBibitem
\bibitem[Satapathy \latin{et~al.}(2022)Satapathy, Liu, Deshmukh, Molinaro,
  Dirnberger, Khatoniar, Koder, and Menon]{satapathy2022thermalization}
Satapathy,~S.; Liu,~B.; Deshmukh,~P.; Molinaro,~P.~M.; Dirnberger,~F.;
  Khatoniar,~M.; Koder,~R.~L.; Menon,~V.~M. Thermalization of Fluorescent
  Protein Exciton--Polaritons at Room Temperature. \emph{Adv. Mater.}
  \textbf{2022}, \emph{34}, 2109107\relax
\mciteBstWouldAddEndPuncttrue
\mciteSetBstMidEndSepPunct{\mcitedefaultmidpunct}
{\mcitedefaultendpunct}{\mcitedefaultseppunct}\relax
\EndOfBibitem
\bibitem[Klimov \latin{et~al.}(1999)Klimov, McBranch, Leatherdale, and
  Bawendi]{klimov1999electron}
Klimov,~V.; McBranch,~D.; Leatherdale,~C.; Bawendi,~M. Electron and hole
  relaxation pathways in semiconductor quantum dots. \emph{Phys. Rev. B}
  \textbf{1999}, \emph{60}, 13740\relax
\mciteBstWouldAddEndPuncttrue
\mciteSetBstMidEndSepPunct{\mcitedefaultmidpunct}
{\mcitedefaultendpunct}{\mcitedefaultseppunct}\relax
\EndOfBibitem
\end{mcitethebibliography}


\providecommand{\latin}[1]{#1}
\makeatletter
\providecommand{\doi}
  {\begingroup\let\do\@makeother\dospecials
  \catcode`\{=1 \catcode`\}=2 \doi@aux}
\providecommand{\doi@aux}[1]{\endgroup\texttt{#1}}
\makeatother
\providecommand*\mcitethebibliography{\thebibliography}
\csname @ifundefined\endcsname{endmcitethebibliography}  {\let\endmcitethebibliography\endthebibliography}{}
\begin{mcitethebibliography}{13}
\providecommand*\natexlab[1]{#1}
\providecommand*\mciteSetBstSublistMode[1]{}
\providecommand*\mciteSetBstMaxWidthForm[2]{}
\providecommand*\mciteBstWouldAddEndPuncttrue
  {\def\EndOfBibitem{\unskip.}}
\providecommand*\mciteBstWouldAddEndPunctfalse
  {\let\EndOfBibitem\relax}
\providecommand*\mciteSetBstMidEndSepPunct[3]{}
\providecommand*\mciteSetBstSublistLabelBeginEnd[3]{}
\providecommand*\EndOfBibitem{}
\mciteSetBstSublistMode{f}
\mciteSetBstMaxWidthForm{subitem}{(\alph{mcitesubitemcount})}
\mciteSetBstSublistLabelBeginEnd
  {\mcitemaxwidthsubitemform\space}
  {\relax}
  {\relax}

\bibitem[Jasrasaria and Rabani(2021)Jasrasaria, and Rabani]{jasrasaria2021interplay}
Jasrasaria,~D.; Rabani,~E. Interplay of surface and interior modes in exciton--phonon coupling at the nanoscale. \emph{Nano Lett.} \textbf{2021}, \emph{21}, 8741--8748\relax
\mciteBstWouldAddEndPuncttrue
\mciteSetBstMidEndSepPunct{\mcitedefaultmidpunct}
{\mcitedefaultendpunct}{\mcitedefaultseppunct}\relax
\EndOfBibitem
\bibitem[Rabani \latin{et~al.}(1999)Rabani, Hetenyi, Berne, and Brus]{rabani1999electronic}
Rabani,~E.; Hetenyi,~B.; Berne,~B.~J.; Brus,~L.~E. Electronic properties of CdSe nanocrystals in the absence and presence of a dielectric medium. \emph{J. Chem. Phys.} \textbf{1999}, \emph{110}, 5355--5369\relax
\mciteBstWouldAddEndPuncttrue
\mciteSetBstMidEndSepPunct{\mcitedefaultmidpunct}
{\mcitedefaultendpunct}{\mcitedefaultseppunct}\relax
\EndOfBibitem
\bibitem[Wall and Neuhauser(1995)Wall, and Neuhauser]{wall1995extraction}
Wall,~M.~R.; Neuhauser,~D. Extraction, through filter-diagonalization, of general quantum eigenvalues or classical normal mode frequencies from a small number of residues or a short-time segment of a signal. I. Theory and application to a quantum-dynamics model. \emph{J. Chem. Phys.} \textbf{1995}, \emph{102}, 8011--8022\relax
\mciteBstWouldAddEndPuncttrue
\mciteSetBstMidEndSepPunct{\mcitedefaultmidpunct}
{\mcitedefaultendpunct}{\mcitedefaultseppunct}\relax
\EndOfBibitem
\bibitem[Toledo and Rabani(2002)Toledo, and Rabani]{toledo2002very}
Toledo,~S.; Rabani,~E. Very large electronic structure calculations using an out-of-core filter-diagonalization method. \emph{J. Comput. Phys.} \textbf{2002}, \emph{180}, 256--269\relax
\mciteBstWouldAddEndPuncttrue
\mciteSetBstMidEndSepPunct{\mcitedefaultmidpunct}
{\mcitedefaultendpunct}{\mcitedefaultseppunct}\relax
\EndOfBibitem
\bibitem[Rohlfing and Louie(2000)Rohlfing, and Louie]{rohlfing2000electron}
Rohlfing,~M.; Louie,~S.~G. Electron-hole excitations and optical spectra from first principles. \emph{Phys. Rev. B} \textbf{2000}, \emph{62}, 4927\relax
\mciteBstWouldAddEndPuncttrue
\mciteSetBstMidEndSepPunct{\mcitedefaultmidpunct}
{\mcitedefaultendpunct}{\mcitedefaultseppunct}\relax
\EndOfBibitem
\bibitem[Philbin and Rabani(2018)Philbin, and Rabani]{philbin2018electron}
Philbin,~J.~P.; Rabani,~E. Electron--hole correlations govern auger recombination in nanostructures. \emph{Nano Lett.} \textbf{2018}, \emph{18}, 7889--7895\relax
\mciteBstWouldAddEndPuncttrue
\mciteSetBstMidEndSepPunct{\mcitedefaultmidpunct}
{\mcitedefaultendpunct}{\mcitedefaultseppunct}\relax
\EndOfBibitem
\bibitem[Jasrasaria \latin{et~al.}(2022)Jasrasaria, Weinberg, Philbin, and Rabani]{jasrasaria2022simulations}
Jasrasaria,~D.; Weinberg,~D.; Philbin,~J.~P.; Rabani,~E. Simulations of nonradiative processes in semiconductor nanocrystals. \emph{J. Chem. Phys.} \textbf{2022}, \emph{157}, 020901\relax
\mciteBstWouldAddEndPuncttrue
\mciteSetBstMidEndSepPunct{\mcitedefaultmidpunct}
{\mcitedefaultendpunct}{\mcitedefaultseppunct}\relax
\EndOfBibitem
\bibitem[Zhou \latin{et~al.}(2013)Zhou, Ward, Martin, Van~Swol, Cruz-Campa, and Zubia]{zhou2013stillinger}
Zhou,~X.; Ward,~D.; Martin,~J.; Van~Swol,~F.; Cruz-Campa,~J.; Zubia,~D. Stillinger-weber potential for the II-VI elements zn-cd-hg-S-se-te. \emph{Phys. Rev. B} \textbf{2013}, \emph{88}, 085309\relax
\mciteBstWouldAddEndPuncttrue
\mciteSetBstMidEndSepPunct{\mcitedefaultmidpunct}
{\mcitedefaultendpunct}{\mcitedefaultseppunct}\relax
\EndOfBibitem
\bibitem[Plimpton(1995)]{plimpton1995fast}
Plimpton,~S. Fast parallel algorithms for short-range molecular dynamics. \emph{J. Comput. Phys.} \textbf{1995}, \emph{117}, 1--19\relax
\mciteBstWouldAddEndPuncttrue
\mciteSetBstMidEndSepPunct{\mcitedefaultmidpunct}
{\mcitedefaultendpunct}{\mcitedefaultseppunct}\relax
\EndOfBibitem
\bibitem[Mandal \latin{et~al.}(2020)Mandal, Krauss, and Huo]{mandal2020polariton}
Mandal,~A.; Krauss,~T.~D.; Huo,~P. Polariton-mediated electron transfer via cavity quantum electrodynamics. \emph{J. Phys. Chem. B} \textbf{2020}, \emph{124}, 6321--6340\relax
\mciteBstWouldAddEndPuncttrue
\mciteSetBstMidEndSepPunct{\mcitedefaultmidpunct}
{\mcitedefaultendpunct}{\mcitedefaultseppunct}\relax
\EndOfBibitem
\bibitem[Mandal \latin{et~al.}(2023)Mandal, Taylor, Weight, Koessler, Li, and Huo]{mandal_theoretical_2023}
Mandal,~A.; Taylor,~M.~A.; Weight,~B.~M.; Koessler,~E.~R.; Li,~X.; Huo,~P. Theoretical Advances in Polariton Chemistry and Molecular Cavity Quantum Electrodynamics. \emph{Chem. Rev.} \textbf{2023}, \emph{123}, 9786--9879\relax
\mciteBstWouldAddEndPuncttrue
\mciteSetBstMidEndSepPunct{\mcitedefaultmidpunct}
{\mcitedefaultendpunct}{\mcitedefaultseppunct}\relax
\EndOfBibitem
\bibitem[Nitzan(2006)]{nitzan2006chemical}
Nitzan,~A. \emph{Chemical dynamics in condensed phases: relaxation, transfer and reactions in condensed molecular systems}; Oxford university press, 2006\relax
\mciteBstWouldAddEndPuncttrue
\mciteSetBstMidEndSepPunct{\mcitedefaultmidpunct}
{\mcitedefaultendpunct}{\mcitedefaultseppunct}\relax
\EndOfBibitem
\end{mcitethebibliography}
\end{document}


\section{Pauli--Fierz Hamiltonian}
We use a single--mode Pauli--Fierz Hamiltonian $H_{\text{PF}}=H_{\text{QD}}+H_{\text{cav}}$ to describe a nanocrystal (NC) quantum dot (QD) (represented by $H_{\text{QD}}$) coupled to the cavity (represented by $H_{\text{cav}}$).
The QD Hamiltonian is given by:\cite{jasrasaria2021interplay}
\begin{equation}
H_{\text{QD}} =E_{g}\left|\psi_g\right\rangle \left\langle \psi_g\right|+\sum_{n}E_{n}\left|\psi_n\right\rangle \langle \psi_n| + \sum_{\alpha}\hbar\omega_{\alpha}b_{\alpha}^{\text{\ensuremath{\dagger}}}b_{\alpha}+\sum_{nm\alpha}V_{nm}^{\alpha}\left|\psi_m\right\rangle \left\langle \psi_n\right|q_{\alpha}.
\label{eq:hamil-excitons}
\end{equation}
In the above equation, $|\psi_g\rangle$ is ground state with energy $E_g$ and $|\psi_n\rangle$ is the $n^{\rm th}$ excitonic state with energy $E_n$. The excitonic states were described as the linear combination of electron--hole pair states:
\begin{equation}
|\psi_n\rangle = \sum_{ai} c_{ai}^{n} {~}|\phi_{a}\rangle \otimes |\phi_{i}\rangle,
\end{equation}
where $a$ and $i$ refer to electron and hole indices, respectively. The single-particle states were calculated using the semi-empirical pseudopotential method,\cite{rabani1999electronic} with parameters fitted to reproduce the band structure and deformation potentials of bulk CdSe.\cite{jasrasaria2021interplay} We used the filter-diagonalization method~\cite{wall1995extraction,toledo2002very} to obtain the electron and hole states near the conduction and valence bands edges, respectively. The coefficients $\left\{ c_{ai}^{n}\right\} $ were obtains by solving the Bethe--Salpeter equation (BSE)~\cite{rohlfing2000electron} with a static dielectric constant $\epsilon=5$,\cite{philbin2018electron} using $140$ hole states and $80$ electron states. More details on the pseudopotential approach can be found in Ref.~\citenum{jasrasaria2022simulations}.

The third term on the right--hand side (RHS) of Eq.~\eqref{eq:hamil-excitons} represents the vibrational modes of the NC. The phonon frequencies $\omega_{\alpha}$ and phonon modes $q_{\alpha}=\sqrt{\frac{\hbar}{2\omega_{\alpha}}}\left(b_{\alpha}^{\text{\ensuremath{\dagger}}}+b_{\alpha}\right)$ were obtained from a force field.
Specifically, we considered a $3.9$~nm CdSe core with $435$ Cd atoms and $435$ Se atoms in the wurtzite crystal structure, with a lattice constant of $a=4.30\thinspace\textup{\AA}$. The Stillinger--Weber force field~\cite{zhou2013stillinger} was used to determine the ground state equilibrium structure and the normal vibrational modes, using LAMMPS.\cite{plimpton1995fast}  The last term on the RHS of Eq.~\eqref{eq:hamil-excitons} represents the exciton-phonon coupling, approximated to first order in the phonon mode coordinates $q_\alpha$. To obtain the exciton phonon coupling matrix, $V_{nm}^{\alpha}$, we first calculated the coupling matrix elements to atom $\nu$ at position vector $R_{\nu k}$ in $k\in\left\{ x,y,z\right\}$ direction:
\begin{equation}
V_{nm}^{\mu k}\equiv \sum_\nu \left\langle \psi_n\left|\left(\frac{\partial v_\nu(\boldsymbol{r}}{\partial R_{\mu k}}\right)_{\boldsymbol{R}_{0}}\right|\psi_m\right\rangle ,
\end{equation}
where $v_\nu(\boldsymbol{r})$ is the pseudopotentials of atom $\nu$. ${\boldsymbol{R}_{0}}$ is the equilibrium configuration of the NC. The exciton-phonon coupling normal mode can then be obtained by rotating $V_{nm}^{\mu k}$ to the normal modes, as further discussed in Ref.~\citenum{jasrasaria2021interplay}.

The cavity Hamiltonian and the cavity-QD coupling are described by the following form:\cite{mandal2020polariton}
\begin{align}
H_{\text{cav}} & =\hbar\omega_{c}a^{\dagger}a+\boldsymbol{k}\cdot\boldsymbol{\mu}\left(\boldsymbol{r}\right)\sqrt{\frac{\hbar\omega_{c}}{2\varepsilon\mathcal{V}}}\left(a^{\dagger}+a\right)+\frac{1}{2\varepsilon\mathcal{V}}\left(\boldsymbol{k}\cdot\boldsymbol{\mu}\left(\boldsymbol{r}\right)\right)^{2}.
\label{eq:hamil_cav}
\end{align}
The first term on the RHS of Eq.~\eqref{eq:hamil_cav} is the single mode photon field, where $a$ ($a^{\dagger}$) is the destruction (creation) operator of a photon mode with frequency $\omega_{c}$. The second and last term on the RHS describes how the exciton transition dipole moment interacts with the photon field. $\boldsymbol{k}$ is a unit vector of the polarization direction of the electromagnetic field, $\varepsilon$ is the effective permittivity inside the cavity, $\mathcal{V}$ is the effective cavity quantization volume, and $\boldsymbol{\mu}\left(\boldsymbol{r}\right)$ represents the total transition dipole operator of the QD system, $\boldsymbol{\mu}\left(\boldsymbol{r}\right)=\sum_{n}\boldsymbol{\mu}_{n}\left(\left|\psi_{g}\right\rangle \left\langle \psi_{n}\right|+\left|\psi_{n}\right\rangle \left\langle \psi_{g}\right|\right)$. 

The last term on the RHS of Eq.~\eqref{eq:hamil_cav} represents the dipole self--energy (DSE). In a simple two--level system with no permanent dipole moment, this term leads to a constant energy shift, and therefore can be ignored, as commonly done in the Jaynes--Cummings (JC) model.\cite{mandal_theoretical_2023}
However, beyond two--level systems, this term can cause state-dependent energy shifts and therefore can play an essential role in polaritonic quantum dynamics. 

Using the relation
\begin{equation}
\hbar g_{n}=\sqrt{\frac{\hbar\omega_{c}}{2\varepsilon\mathcal{V}}}\boldsymbol{\mu}_{n}\cdot\boldsymbol{k},
\end{equation}
the cavity Hamiltonian can be written as:
\begin{equation}
H_{\text{cav}}=\hbar\omega_{c}a^{\dagger}a+\sum_{n}\hbar g_{n}\left(\left|\psi_{g}\right\rangle \left\langle \psi_{n}\right|+\left|\psi_{n}\right\rangle \left\langle \psi_{g}\right|\right)\left(a^{\dagger}+a\right)+H_{\text{DSE}},
\end{equation}
where
\begin{align}
H_{\text{DSE}} &=\frac{1}{2}\frac{1}{\varepsilon\mathcal{V}}\left(\boldsymbol{k}\cdot\boldsymbol{\mu}\right)^{2}=\frac{\hbar}{\omega_{c}}\left(\sum_{n}g_{n}\left(\left|\psi_{g}\right\rangle \left\langle \psi_{n}\right|+\left|\psi_{n}\right\rangle \left\langle \psi_{g}\right|\right)\right)^{2}\nonumber \\
 & =\frac{\hbar}{\omega_{c}}\left[\text{\ensuremath{\sum_{n}g_{n}^{2}}}\left|\psi_{g}\right\rangle \left\langle \psi_{g}\right|+\sum_{nm}g_{n}g_{m}\left|\psi_{n}\right\rangle \left\langle \psi_{m}\right|\right].
\end{align}
If we focus on in single--excitation manifold, we can drop the counter--rotating term $\sum_{n}\hbar g_{n}(|\psi_{g}\rangle \langle \psi_{n}|a+|\psi_{n}\rangle \langle \psi_{g}|a^{\dagger})$.

\section{Polaron transformation}
The Pauli--Fierz Hamiltonian in the polaritonic representation, $\left|\varphi_{n}\right\rangle$, is given by:
\begin{equation}
\tilde{H}_{\text{PF}} =\sum_{n}\tilde{E}_{n}\left|\varphi_{n}\right\rangle \left\langle \varphi_{n}\right|+\sum_{\alpha}\hbar\omega_{\alpha}b_{\alpha}^{\text{\ensuremath{\dagger}}}b_{\alpha}
+  \sum_{mn\alpha}\tilde{V}_{mn}^{\alpha}\left|\varphi_{m}\right\rangle \left\langle \varphi_{n}\right|q_{\alpha},
\label{eq:hamil-total}
\end{equation}
where $\left|\varphi_{n}\right\rangle =c_{ng}\left|\psi_g,1\right\rangle +\sum_{m}c_{nm}\left|\psi_{m},0\right\rangle $ is a polaritonic state with energy $\tilde{E}_{n}$ (obtained by finding the eigenvalues of $H_\text{S}$ (see Eq.~(3) in the main text)). $\tilde{V}_{mn}^{\alpha}$ represents the coupling matrix element between two polaritonic states $\left|\varphi_{m}\right\rangle $ and $\left|\varphi_{n}\right\rangle $ via phonon mode $\alpha$, and is obtained by applying the same unitary transformation on $H_\text{S}$ to ${V}_{mn}^{\alpha}$ for each mode $\alpha$.

The polaron transformation of $H_\text{PF}$ is similar to the one developed for excitons,\cite{jasrasaria2023circumventing} namely, ${\cal H}=e^{S}\tilde{H}_{\text{PF}}e^{-S}$,
where $S=-\sum_{\alpha}\frac{i\sum_{n}\tilde{V}_{nn}^{\alpha}p_{\alpha}}{\hbar\omega_{\alpha}^{2}}\left|\varphi_{n}\right\rangle \left\langle \varphi_{n}\right|$
and $p_{\alpha}$ is the momentum operator of vibrational mode $\alpha$.
The total transformed Hamiltonian is then given by:\cite{jasrasaria2023circumventing} 
\begin{equation}
{\cal H}  =\sum_{n}\left(\tilde{E}_{n}-\lambda_{n}\right)\left|\varphi_{n}\right\rangle \left\langle \varphi_{n}\right|+\sum_{\alpha}\hbar\omega_{\alpha}b_{\alpha}^{\text{\ensuremath{\dagger}}}b_{\alpha} 
+ \sum_{n\neq m}\left(\sum_{\alpha}W_{nm}^{\alpha}q_{\alpha}-\lambda_{nm}\right)\left|\varphi_{n}\right\rangle \left\langle \varphi_{m}\right|\label{eq:Hamil-total-polaron},
\end{equation}
where $\lambda_{n}=\frac{1}{2}\sum_{\alpha}\left(\tilde{V}_{nn}^{\alpha}\right)^{2}/\omega_{\alpha}^{2}$
is the reorganization energy (polaron shift) of the polaritonic state
$\left|\varphi_{n}\right\rangle $, $\lambda_{nm}=\frac{1}{2}\sum_{\alpha}W_{nm}^{\alpha}\left(\tilde{V}_{mm}^{\alpha}+\tilde{V}_{nn}^{\alpha}\right)/\omega_{\alpha}^{2}$,
and the rescaled couplings between the polaritonic states and the vibrational mode $\alpha$ is given by $W=e^{S}\tilde{V}e^{-S}$,
with matrix elements:
\begin{equation}
W_{nm}^{\alpha}  =\exp\left(-\frac{i}{\hbar}\sum_{\gamma}\frac{p_{\gamma}\tilde{V}_{nn}^{\gamma}}{\omega_{\gamma}^{2}}\right)
 \tilde{V}_{nm}^{\alpha}\exp\left(+\frac{i}{\hbar}\sum_{\gamma}\frac{p_{\gamma}\tilde{V}_{mm}^{\gamma}}{\omega_{\gamma}^{2}}\right).
\end{equation}
We use the definition:
${\cal
H}=\mathcal{H}_{\text{S}}+\mathcal{H}_{\text{B}}+\mathcal{H}_{\text{I}}$, 
where
\begin{equation}
\mathcal{H}_{\text{S}}=\sum_{n}\left(\tilde{E}_{n}-\lambda_{n}\right)\left|\varphi_{n}\right\rangle \left\langle \varphi_{n}\right|
\end{equation}
\begin{equation}
\mathcal{H}_{\text{B}}=\sum_{\alpha}\hbar\omega_{\alpha}b_{\alpha}^{\text{\ensuremath{\dagger}}}b_{\alpha}
\end{equation}
\begin{equation}
\mathcal{H}_{\text{I}} =\sum_{n\neq m}\left(\sum_{\alpha}W_{nm}^{\alpha}q_{\alpha}-\lambda_{nm}\right)\left|\varphi_{n}\right\rangle \left\langle \varphi_{m}\right|.
\end{equation}


\section{Non--Markovian local--time secular Redfield equation}
We use the time-local Redfield equations to solve for the reduced density matrix in the polaritonic basis $\sigma(t)$:\cite{nitzan2006chemical}
\begin{align}
(n\neq m):\frac{d\sigma_{nm}}{dt} & =-i\omega_{nm}\sigma_{nm}\left(t\right)-\sum_{c}R_{nl,ln}\left(\omega_{nl},t\right)\sigma_{nm}\left(t\right)-\sum_{c}R_{ml,lm}^{*}\left(\omega_{ml},t\right)\sigma_{nm}\left(t\right)\\
(n=m):\frac{d\sigma_{nn}}{dt} & =-\sum_{c}R_{nl,ln}\left(\omega_{nl},t\right)\sigma_{nn}\left(t\right)+\sum_{c}R_{ln,nl}\left(\omega_{ln},t\right)\sigma_{ll}\left(t\right)\nonumber \\
 & +\sum_{c}R_{ln,nl}^{*}\left(\omega_{ln},t\right)\sigma_{ll}\left(t\right)-\sum_{c}R_{nl,ln}^{*}\left(\omega_{nl},t\right)\sigma_{nn}\left(t\right),
\end{align}
where the time--dependent Redfield tensor $R_{nm,kl}$ is written as:
\begin{equation}
R_{nm,kl}\left(t\right)=\int_{0}^{t}d\tau M_{nm,kl}\left(\tau\right)e^{i\omega_{lk}\tau}
\end{equation}
\begin{equation}
\hbar\omega_{nm}=\delta E_{nm}=\left(\tilde{E}_{n}-\lambda_{n}\right)-\left(\tilde{E}_{m}-\lambda_{m}\right).
\end{equation}
$M_{nm,kl}$ is the polaron--transformed system--bath interaction time correlation function:
\begin{equation}
M_{nm,kl}\left(t\right)=\left\langle \left(\sum_{\alpha}W_{nm}^{\alpha}\left(t\right)q_{\alpha}\left(t\right)-\lambda_{nm}\left(t\right)\right)\left(\sum_{\alpha}W_{kl}^{\alpha}\left(0\right)q_{\alpha}\left(0\right)-\lambda_{kl}\left(0\right)\right)\right\rangle _{\text{B}}/\hbar^{2},
\end{equation}
where $\langle \cdots\rangle_{\text{B}}$  is the phonon bath equilibrium thermal average, $q_{\alpha}(t)=\sqrt{\frac{\hbar}{2\omega_{\alpha}}}\left(b_{\alpha}^{\text{\ensuremath{\dagger}}}e^{i\omega_{\alpha}t}+b_{\alpha}e^{-i\omega_{\alpha}t}\right)$,
$p_{\alpha}(t)=i\sqrt{\frac{\hbar\omega_\alpha}{2}}\left(b_{\alpha}^{\text{\ensuremath{\dagger}}}e^{i\omega_{\alpha}t}-b_{\alpha}e^{-i\omega_{\alpha}t}\right)$, 
$W_{nm}^{\alpha}\left(t\right)=\exp\left(-\frac{i}{\hbar}\sum_{\gamma}\frac{p_{\gamma}(t)\tilde{V}_{nn}^{\gamma}}{\omega_{\gamma}^{2}}\right)\tilde{V}_{nm}^{\alpha}\exp\left(+\frac{i}{\hbar}\sum_{\gamma}\frac{p_{\gamma}(t)\tilde{V}_{mm}^{\gamma}}{\omega_{\gamma}^{2}}\right)$, 
and $\lambda_{mn}(t)=\frac{1}{2}\sum_{\alpha}W_{nm}^{\alpha}(t)\left(\tilde{V}_{mm}^{\alpha}+\tilde{V}_{nn}^{\alpha}\right)/\omega_{\alpha}^{2}$.

The transition rates between two eigenstates $\left|\varphi_{n}\right\rangle $ and $\left|\varphi_{m}\right\rangle $:
\begin{align}
\varGamma_{n\rightarrow m}\left(t\right) & =R_{nm,mn}\left(\omega_{nm},t\right)+R_{nm,mn}^{*}\left(\omega_{nm},t\right)\nonumber \\
 & =\int_{0}^{t}d\tau M_{nm,mn}\left(\tau\right)e^{i\omega_{nm}\tau}+\int_{0}^{t}d\tau M_{nm,mn}\left(\tau\right)e^{-i\omega_{nm}\tau}\nonumber \\
 & =\int_{-t}^{t}d\tau M_{nm,mn}\left(\tau\right)e^{i\omega_{nm}\tau}.
\end{align}
At enough longer time, $\varGamma_{n\rightarrow m}\left(t\right)$ becomes independent of time and is the Fourier transformation of correlation function $M_{nm,mn}\left(\tau\right)$ at energy gap $\hbar\omega_{nm}$.

\section{Polariton dynamics}
\begin{figure}[H]
\begin{centering}
\includegraphics[width=17.5cm]{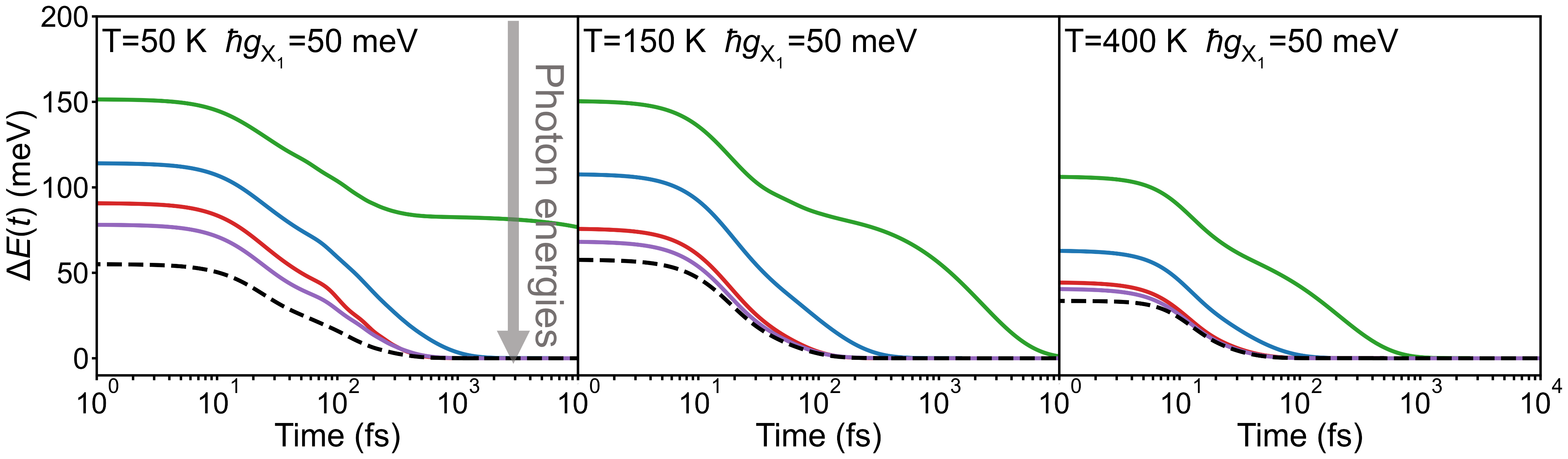}
\par\end{centering}
\caption{The energy relaxation, $\Delta E\left(t\right)=\left\langle H_{\text{S}}\right\rangle \left(t\right)-\left\langle H_{\text{S}}\right\rangle _{{\rm eq}}$, for $50$K (panel (a)), $150$K (panel (b)), and $400$K (panel (c)). The photon energy is set to be $\hbar\omega_{c}=2.02$, $2.08$, $2.14$, $2.20$~eV. The cavity--exciton coupling strength $\hbar g_{\text{X}_{1}}=50\thinspace\text{meV}$. Color codes are similar to Fig.~4 in the main text.}
\label{fig:sys_relax_SI}
\end{figure}

In this subsection, we show additional results for the population dynamics and the energy relaxation for different temperatures not presented in the main text. The population dynamics for a NC placed in an optical cavity at $T=50\text{\,K}$, $150\text{\,K}$, and $400\text{\,K}$ are shown in Fig.~\ref{fig:sys_relax_SI}), panels (a)-(c), respectively. Similarly to the room temperature results shown in Fig.~3, as the photon energy decreases below the onset of absorption, the relaxation slow down considerably. This is much more pronounced at low temperatures (see panel (a) for $T=50$ K), where we find that the energy of the system does not relax to its equilibrium value even at very long times.

\begin{figure}[H]
\begin{centering}
\includegraphics[width=15cm]{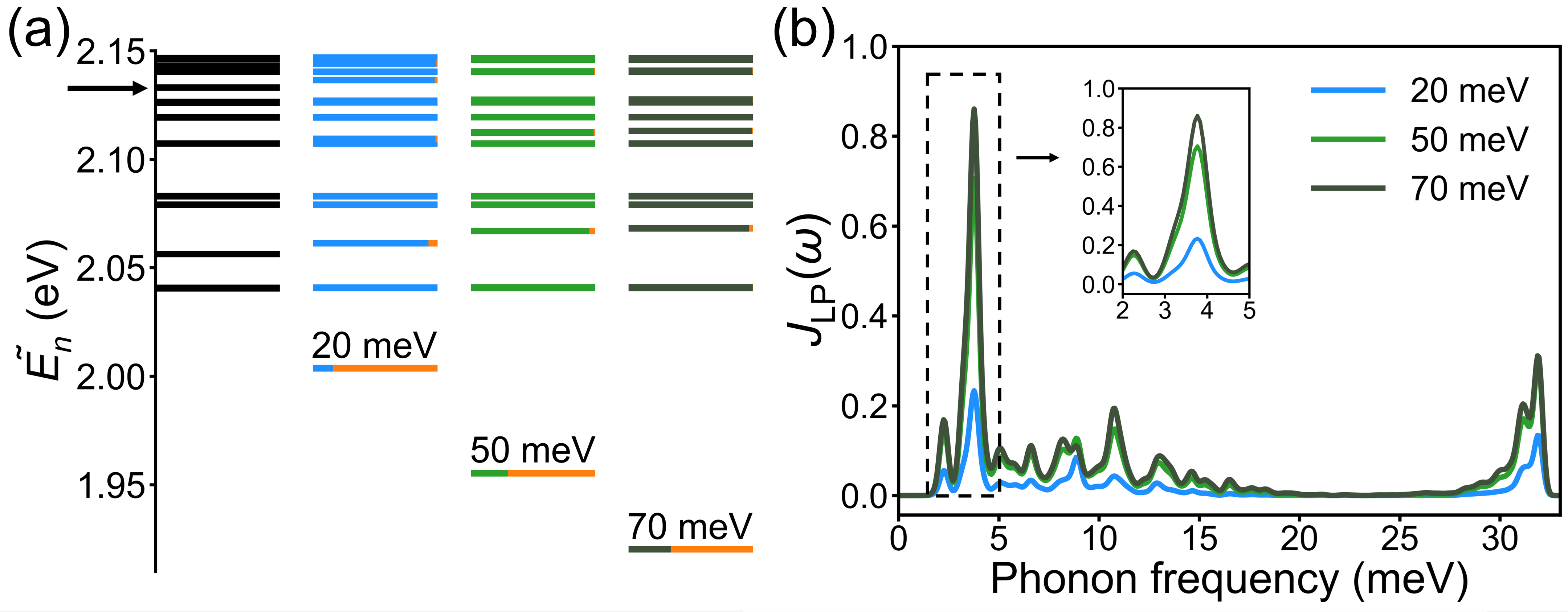}
\par\end{centering}
\caption{(a) Polaritonic or excitonic energies for different cavity--exciton coupling strength for $\hbar\omega_{c}=2.02\thinspace\text{eV}$, $\hbar g_{\rm{X}_{1}}=20$, $50$, and $70$~meV. The photon fraction is depicted by the orange color for each polaritonic state. The black arrow on the left side of the panel is the initial excitonic state. (b) The spectral density computed for the lowest polaritonic state as a function of phonon frequency for different cavity--exciton coupling strengths. We use the same color code as in Fig.~4 of the main text.  Insets: Zoom over the acoustic frequency range.}
\label{fig:DOS_SI}
\end{figure}

To explain the dependence on the cavity coupling strength, we analyze the level structure and spectral densities in Fig.~\ref{fig:DOS_SI}. This plot is similar to Fig.~4 in the main text, but instead of varying the photon energy, Fig.~\ref{fig:DOS_SI} shows the results for different coupling strengths.  Panel (a) shows the polaritonic energies ($\tilde{E_{n}}$) in the absence of a cavity (excitonic energies, $E_{n}$) and for $3$ different cavity--exciton  coupling strengths. The cavity photon energy is set to $\hbar \omega_{c} =2.02 $~eV. In panel (b), we plot the spectral density, $J_{\text{LP}}\left(\omega\right)=\frac{1}{2}\sum_{\alpha}\left(\frac{\tilde{V}_{\text{LP}}^{\alpha}}{\omega_{\alpha}}\right)^{2}\delta\left(\omega-\omega_{\alpha}\right)$, for the LP for different values of cavity--exciton coupling strengths.
As the coupling to the cavity increases, the photon fraction of the LP state decreases and the overall polariton--phonon coupling increases. This enhanced vibronic coupling competes with the enlarged gap, but only provides trivial counterwork to the exponential suppression brought by the enlarged gap. A bottleneck phenomenon is still here for a large cavity--exciton coupling strength.

\section{Analysis of the transition rate}
\subsection{Short-time behavior of $M_{nm,mn}\left(t\right)$}
\begin{figure}[H]
\begin{centering}
\includegraphics[width=8cm]{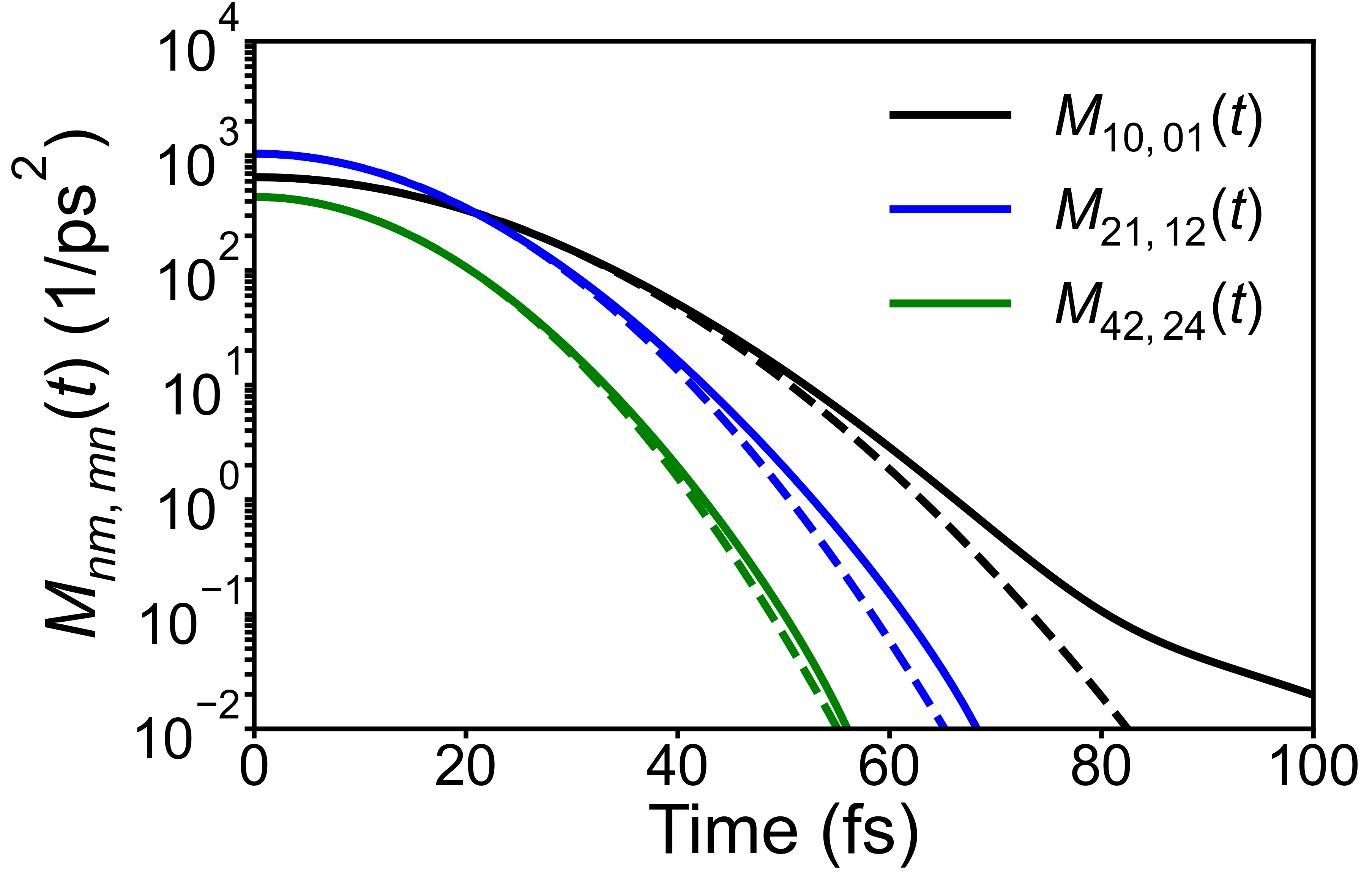}
\par\end{centering}
\caption{Plot of some representative $M_{nm,mn}(t)$ in absence of cavity. The solid lines are in exact form, the dash lines are in Gaussian form. }  
\label{fig:corr_func}
\end{figure}

In Fig.~\ref{fig:corr_func} we plots $M_{nm,mn}\left(t\right)$ for representative elements. It is clear that $M_{nm,mn}\left(t\right)$ decays rapidly in time and can be approximated by a Gaussian form over several orders of decay. The Gaussian form can be obtained from a short time expansion of $M_{nm,mn}\left(t\right) = M_{nm,mn}\left(0\right) e^{-\eta_{nm} (T) t^{2} / \hbar^{2}}$, where
\begin{equation}
    \eta_{nm}\left(T\right)  = -\frac{\hbar^{2}}{\Lambda_{nm}}\sum_{\alpha}\left|\tilde{V}_{nm}^{\alpha}\right|^{2}+2k_{{\rm B}}T\left(\lambda_{n}+\lambda_{m}-\sum_{\alpha}\frac{1}{\omega_{\alpha}^{2}}\tilde{V}_{nn}^{\alpha}\tilde{V}_{mm}^{\alpha}\right).
\end{equation}
and
\begin{equation}
M_{nm,mn}\left(0\right) \approx\sum_{\alpha\beta}\tilde{V}_{n,m}^{\alpha}\tilde{V}_{m,n}^{\beta}\left\langle q_{\alpha}q_{\beta}\right\rangle /\hbar^{2}=\frac{k_{\text{B}}T}{\hbar^{2}}\Lambda_{nm}.
\end{equation}
with $\Lambda_{nm} =\sum_{\alpha} \left| \tilde{V}_{nm}^{\alpha} \right|^{2}/ \omega_{\alpha}^{2}$. The rate of transition between the two polaritonic states with transition gap $\delta E_{nm}$ is given by the Fourier transform of the above: 
\begin{equation}
\varGamma_{n\rightarrow m}(T) = \mathcal{F}\left[M_{nm,mn}\left(t\right)\right]=M_{nm,mn}\left(0\right)\sqrt{\frac{\hbar^{2}\pi}{\eta_{nm}(T)}}e^{-\delta E_{nm}^{2}/4\eta_{nm}(T)}.
\end{equation}
This is the expression we use in the main text to analyze the transition in the polaritonic bottleneck.

\subsection{Bottleneck regime}
\begin{figure}[H]
\begin{centering}
\includegraphics[width=8cm]{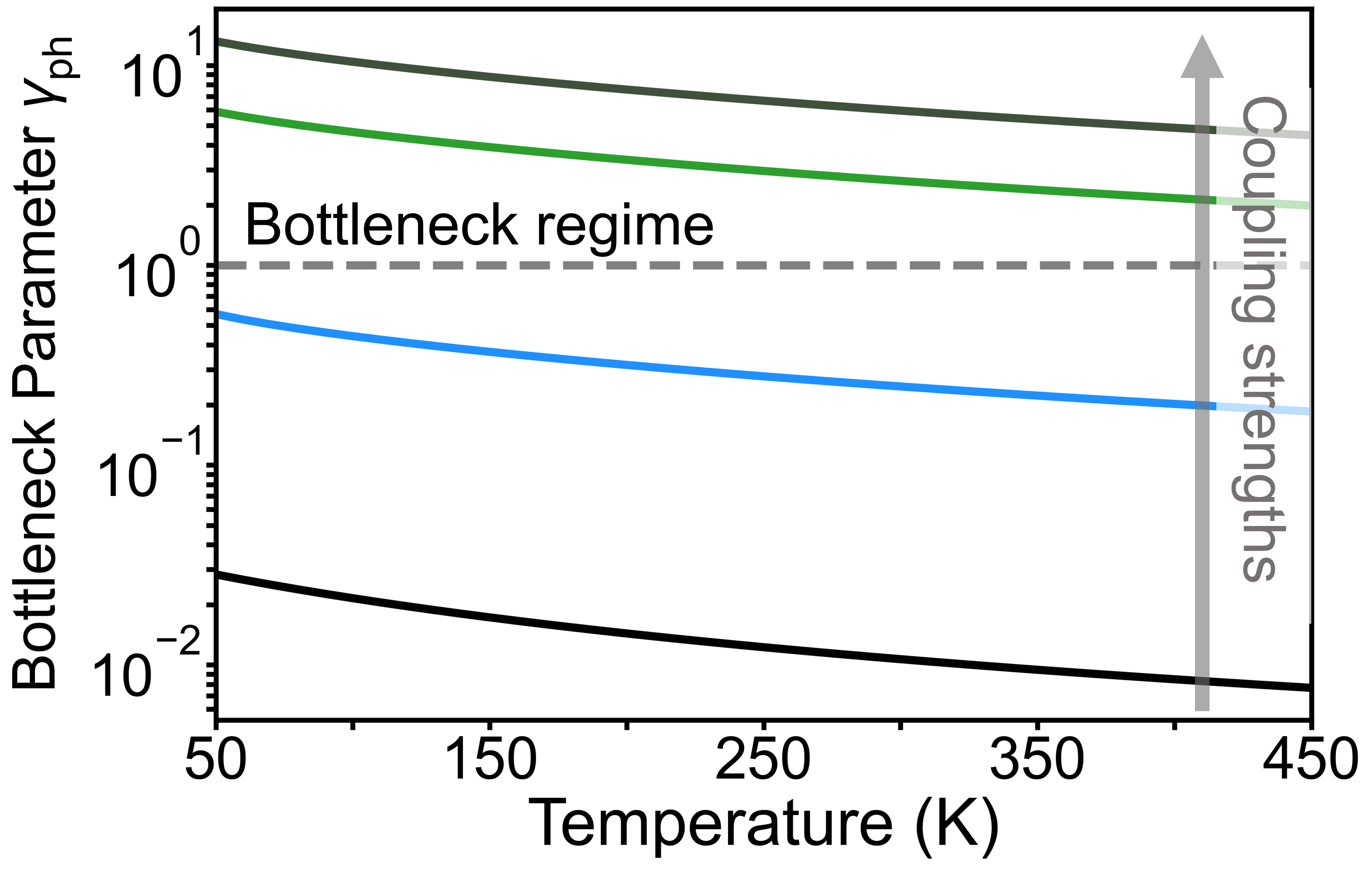}
\par\end{centering}
\caption{Plot of $\gamma_{\rm ph}$ vs. $T$ for $\hbar\omega_{c}=2.02$~eV and for $\hbar g_{\text{X}_{1}} = 0,$ $20,$ $50,$ $70$~meV. The gray dash line is for $\gamma_{\rm ph}=1$. 
\label{fig:eta_SI}}
\end{figure}

In Fig.~\ref{fig:eta_SI}, we plot bottleneck parameter $\gamma_{\rm ph}$ as a function of temperature (similar to Fig.~5 in the main text) for a given photon energy of $2.02\thinspace\text{eV}$ for different values of the cavity-exciton coupling strengths (rather than for different photon energies). As can be seen, the behavior with increasing coupling strength is similar to that observed when the photon energy decreases below the onset of absorption. 
\bibliography{polariton}